\documentclass{aa}
\usepackage{array}
\newcolumntype{P}[1]{>{\centering\arraybackslash}p{#1}}
\usepackage{txfonts}
\usepackage{epsfig}
\usepackage{rotating}
\usepackage{times}
\usepackage{booktabs}
\usepackage{longtable}
\usepackage{color}
\usepackage{natbib}
\usepackage{amsmath}
\usepackage[utf8]{inputenc}
\usepackage{caption}
\usepackage{float}
\usepackage{pdflscape}

\newcommand{\HeI} {He\,{\sc i}}
\newcommand{\HeII}{He\,{\sc ii}}

\newcommand{\CIV}{C\,{\sc iv}}

\newcommand\NV{N\,{\sc v}}

\newcommand\AlIII{Al\,{\sc iii}}

\newcommand\SiIII{Si\,{\sc iii}}
\newcommand\SiIV{Si\,{\sc iv}}

\newcommand{\FeIII}{Fe\,{\sc iii}}

\bibpunct{(}{)}{;}{a}{}{,}

\begin{document}

   \title{Measuring the stellar wind parameters in IGR~J17544-2619 and Vela~X-1 constrains the accretion physics in Supergiant Fast X-ray Transient and classical Supergiant X-ray Binaries }

\author{A. Gim\'{e}nez-Garc\'{i}a\inst{1,3}
\and
T. Shenar\inst{2}
\and
J.~M.~Torrej\'{o}n\inst{1,3}
\and
L. Oskinova\inst{2}
\and
S. Mart\'{i}nez-N\'{u}\~{n}ez\inst{1}
\and
W.-R. Hamann\inst{2}
\and
J. J. Rodes-Roca\inst{1,3,5}
\and
A. Gonz\'{a}lez-Gal\'{a}n\inst{2}
\and
J.~Alonso-Santiago\inst{1}
\and
C. Gonz\'{a}lez-Fern\'{a}ndez\inst{4}
\and
G. Bernabeu\inst{1,3}
\and
A.~Sander\inst{2}
}

   \authorrunning{A. Gim\'{e}nez-Garc\'{i}a\inst{1}}
   \titlerunning{Title}

   \offprints{A. Gim\'{e}nez-Garc\'{i}a}

\institute{Departamento de F\'{\i}sica, Ingenier\'{\i}a de Sistemas y 
Teor\'{\i}a de la Se\~{n}al, 
University of Alicante, P.O. Box 99, E03080 Alicante, Spain. 
\email{angelgimenez@ua.es}
\and
Institut f\"ur Physik und Astronomie, Universit\"at Potsdam,
Karl-Liebknecht-Str. 24/25, D-14476 Potsdam, Germany
\and Instituto Universitario de F\'isica Aplicada a las Ciencias y las 
Tecnolog\'ias, University of Alicante, 
P.O. Box 99, E03080 Alicante, Spain
\and
Institute of Astronomy, University of Cambridge, Madingly
Road, Cambridge, CB3 0HA, UK
\and
MAXI team, Institute of Physical and Chemical Research (RIKEN), 2-1 Hirosawa, 
Wako, Saitama, 351-0198, Japan
}

   \date{Received date; accepted date}

 
  \abstract
   {Classical Supergiant X-ray Binaries (SGXBs) and Supergiant 
Fast X-ray Transients (SFXTs) are two types of High-mass X-ray Binaries (HMXBs) that 
present similar donors but, at the same time, show very different behavior in 
the X-rays. The reason for this dichotomy of wind-fed HMXBs is still a matter of debate. 
Among the several explanations that have been proposed, some of them invoke specific 
stellar wind properties of the donor stars. Only dedicated empiric analysis
of the donors' stellar wind can provide the required information to accomplish an
adequate test of these theories. However, such analyses are scarce.}
   {To close this gap, we perform a comparative analysis of the optical companion
in two important systems: IGR~J17544-2619 (SFXT) and Vela~X-1 (SGXB). 
We analyse the spectra of each star in detail and derive their stellar and wind properties. 
As a next step, we compare the wind parameters, giving us an excellent chance 
of recognizing key differences between donor winds in SFXTs and SGXBs. }
   {We use archival infrared, optical and ultraviolet observations, and analyse them 
with the non-LTE Potsdam Wolf-Rayet model atmosphere code. We derive the physical 
properties of the stars and their stellar winds, accounting for the influence of X-rays 
on the stellar winds.}
  {We find that the stellar parameters derived from the analysis 
generally agree well with the spectral types of the two donors: O9I (IGR~J17544-2619) 
and B0.5Iae (Vela~X-1). The distance to the sources have been revised and also agrees
well with the estimations already available in the literature. In IGR~J17544-2619 we are able to narrow
the uncertainty to $\text{d} = 3.0 \pm 0.2$~kpc. From the stellar radius of the donor and its X-ray behavior, the
eccentricity of IGR~J17544-2619 is constrained to $e<0.25$. The derived chemical abundances
point to certain mixing during the lifetime of the donors. An important difference between the
stellar winds of the two stars is their terminal velocities ($\varv_{\infty}=1500$~km/s in IGR~J17544-2619
and $\varv_{\infty}=700$~km/s in Vela~X-1), which has important consequences on the 
X-ray luminosity of these sources.   }
   {The donors of IGR~J17544-2619 and Vela~X-1 have similar spectral types 
as well as similar parameters that physically characterise them and their spectra. In 
addition, the orbital parameters of the systems are similar too, with a nearly circular 
orbit and short orbital period. However, they show moderate differences in their stellar wind
velocity and spin period of their neutron stars that have a strong impact on the X-ray luminosity
of the sources. This specific combination of wind speed and pulsar spin favours an accretion regime
with a persistently high luminosity in Vela~X-1, while it favours an inhibiting accretion mechanism 
in IGR~J17544-2619. Our study demonstrates that the wind relative velocity is critical 
in the determination of the class of HMXBs hosting a supergiant donor, given that it may shift
the accretion mechanism from direct accretion to propeller regimes when combined with other parameters.  }


   \maketitle

\parindent=0pt

\setlongtables

\section{Introduction}
Within the wide zoo of High-mass X-ray Binaries~(HMXBs), we find two classes of 
sources where a compact object, usually a neutron star, accretes matter from the 
stellar wind of a supergiant OB donor. These are the classical Supergiant 
X-ray Binaries~(SGXBs) and the Supergiant Fast X-ray Transients~(SFXTs). These 
two groups of systems, despite hosting roughly the same type of stars, have 
distinctive properties when observed in the X-rays.  \\

Supergiant X-ray Binaries are persistent sources, with an X-ray luminosity in the range 
$L_\text{X} \sim 10^{33-39}$~erg/s. They are often variable, showing flares and off-states 
that indicate abrupt changes in the accretion rate 
\citep{2008A&A...492..511K, 2014A&A...563A..70M}. However, their variability is 
not as extreme as in SFXTs \citep{2007A&A...476..335W}. 
The dynamic range (ratio between luminosity in 
outburst and in quiescence) in SGXBs is $\lesssim 2$ orders of magnitude. In 
contrast, the dynamic range in SFXTs can reach up to six orders of magnitude in 
the most extreme cases such as IGR~J17544-2619 \citep{2015A&A...576L...4R, 2005A&A...441L...1I}, 
analysed in this work. During 
quiescence, SFXTs exhibit a low X-ray luminosity of $L_\text{X} \sim 10^{32}$~erg/s 
\citep{2005A&A...441L...1I}, but 
they spend most of their time in an intermediate level of emission of $\sim 
10^{33-34}$~erg/s \citep{2008ApJ...687.1230S}. They display short outbursts ($\sim$few hours), reaching 
luminosities up to $10^{36-37}$ erg/s \citep{2011arXiv1111.5747S, 
2009ApJ...690..120S}. \\

There are other sources in between SGXBs and SFXTs, the so called "intermediate SFXTs", 
which have a dynamic range of $\gtrsim 2$ orders of magnitude. Hence, there is no sharp border 
that clearly separates SGXBs and SFXTs. The categorization of SFXTs as a new class of HMXBs
\citep{2006ESASP.604..165N} was possible thanks to \textit{INTEGRAL} observations 
\citep{2005A&A...444..221S}. Since then, several explanations have been proposed 
in order to explain their transient behavior. \\

\cite{2008AIPC.1010..252N} suggested that the intrinsic clumpiness of the wind 
of hot supergiant donors, together with different orbital configurations, may 
explain the different dynamic ranges between SGXBs and SFXTs. If the 
eccentricity of SFXTs is high enough, the compact object swings between dense 
regions with a high probability of accreting a wind clump and flare up, and 
diffuse regions where this probability is low and the source is consequently 
faint in the X-rays. In SGXBs, the compact object would orbit in a closer and 
more circular trajectory, accreting matter incessantly. However, the short orbital period of some SFXTs is contradictory with this scheme \citep{2015A&ARv..23....2W}. \\

Other ingredients, such as the magnetic field of the neutron star and/or the 
the spin period, might be important. This is supported by the monitoring of 
SFXTs. Tracing SFXTs for a long period, \citet{2013MNRAS.431..327L} conclude 
that, in SFXTs, the accretion is notably inhibited most of the time. 
One can invoke to the different possible configurations of 
accretion, co-rotation and magnetospheric radius in order to relax the extremely 
sharp density contrast required in the above mentioned interpretation \citep{2007AstL...33..149G, 2008ApJ...683.1031B, 2010arXiv1004.0293G}. 
The size of these radii depend on the wind, orbital, and neutron star parameters. For 
instance, if the magnetospheric radius is larger than the accretion radius 
\citep{1952MNRAS.112..195B}, the inflow of matter is significantly inhibited by 
a magnetic barrier, resulting in a relatively low X-ray emission from the 
source. Under this interpretation, the physical 
conditions in SFXTs make them prone to regime transitions as a response to 
relatively modest variations in the wind properties of the donor, which cause 
abrupt changes in X-ray luminosity. \\

These changes might also be explained within the theory of quasi-spherical accretion 
onto slowly rotating magnetized neutron 
stars developed by \cite{2012MNRAS.420..216S}. 
This theory describes the so-called 
\textit{subsonic settling accretion regime} in detail. In slowly rotating 
neutron stars, the penetration of matter into the magnetosphere is driven 
predominantly by Rayleigh-Taylor instabilities \citep{1976Natur.262..356E}. When 
the cooling of the plasma in the boundary of the magnetosphere is not 
sufficiently efficient, the accretion of matter is highly inhibited and 
consequently the X-ray luminosity is low. On the other hand, when the cooling 
time is much smaller than the characteristic free-fall time ($t_{\rm cool} \ll 
t_{\rm ff}$), the instability conditions are fulfilled and the plasma easily 
enters the magnetosphere, triggering high X-ray luminosity. The last is achieved 
when the X-ray luminosity is $L_\text{X} \gtrsim 4 \cdot 10^{36}$~erg/s, and the rapid 
Compton cooling dominates over the radiative cooling. For the brightest flares 
($L_\text{X} > 10^{36}$), \cite{2014MNRAS.442.2325S} proposed that a magnetized wind of 
the donor might induce magnetic reconnection, enhancing the accretion up to the 
critical X-ray luminosity and triggering the suction of the whole shell by the 
neutron star. \\

We need as much information as possible about the stellar wind conditions in 
order to understand the different behavior of SGXBs and SFXTs. However, very few 
analyses of SGXBs and SFXTs have been performed so far in the 
ultraviolet-optical-infrared spectral range  using modern atmosphere codes which 
include NLTE and line blanketing effects. Moreover, although the X-rays are 
mainly produced in the surroundings of the compact object, the analysis of 
X-rays observations is directly affected by the physical properties of the donor 
and its wind. For instance, the assumed abundances strongly affect the derived 
value of one of the most important parameters in the X-rays studies: the 
equivalent hydrogen column density ($\text{N}_\text{H}$). More reliable abundances make the 
$\text{N}_\text{H}$ estimations more reliable. Analysing spectra by means of line-blanked, 
NLTE model atmosphere codes is currently the best way to extract the stellar 
parameters of hot stars with winds. \\

In this work we analyze the optical companion of two X-ray sources: 
IGR~J17544-2619~(SFXT) and Vela~X-1~(SGXB). These sources are usually considered 
to be \textit{prototypical} for their respective classes  
\citep{2014A&A...563A..70M, 2009ApJ...690..120S, 2007PThPS.169..196M}. Hence, in 
addition to the important scientific value of studying these sources by 
themselves, this is an excellent opportunity to compare the donor's parameters 
in these two prototypical systems, and to test how well the aforementioned 
resolutions for the SFXT puzzle fit in with our results. \\

The structure of the paper is as follows. In Sect.~\ref{sec:obs} we describe 
the set of observations used in this work. In Sect.~\ref{sec:PoWR} we explain 
the main features of Potsdam Wolf-Rayet~(PoWR) code employed in the fits. In 
Sect.~\ref{sec:fit} we detail the fit process and give the obtained results. 
In Sect.~\ref{sec:disc} we discuss several consequences arising from our 
results. Finally, in Sect.~\ref{sec:conclusions} we enumerate the conclusions 
that we find from this work. 

\section{The observations \label{sec:obs}}
In this study we used data from International Ultraviolet 
Explorer~(IUE)\footnote{available at https://archive.stsci.edu/iue/}, the 
fiber-fed extended range optical spectrograph~(FEROS)\footnote{available at 
http://archive.eso.org/} operated at the European Southern Observatory~(ESO) in 
La~Silla, Chile; and the infrared~(IR) spectrograph SpeX in the NASA Infrared 
Telescope Facility~(IRTF) in Mauna~Kea, Hawaii. \\

The IUE is provided with two spectrographs (long-wavelength in the range $1850-3300$~\AA ~and short-
wavelength in $1150-2000$~\AA) and four cameras (prime and redundant camera, for each spectrograph). Each
spectrograph can be used with either large aperture (a slot 10x20~arcsec), or small aperture (a circle
$3$~arcsec diameter). In addition, each spectrograph has two dispersion modes: high resolution and low
resolution. High resolution mode ($\sim 0.2$~\AA) utilizes an echelle grating plus a cross-disperser. Low
resolution mode ($\sim 6$~\AA) utilizes only the cross-disperser. IUE provides flux calibrated data. This
is an important advantage due to two main reasons: first, we used these observations to fit the spectral
energy distribution from the models, as explained below in Sect.~\ref{sec:vela}; and second, we did not
have to normalize the UV spectrum. As we can see in Fig.~\ref{fig:summ_UVVela} and \ref{fig:SED_Vela},
it is not straightforward to see the actual flux level of the UV continuum, since this spectral range is
almost completely covered by spectral lines. Therefore, any normalization by visual inspection would lead
to significant errors. Instead, we rectified the IUE spectra using the 
PoWR model continuum. \\

FEROS is a spectrograph that yields high resolution echelle spectroscopy ($R \sim 48000$) and
high efficiency ($\sim 20 \%$) in the optical wavelength range ($3600-9200$~\AA)
\citep{1999Msngr..95....8K}. SpeX is an infrared spectrograph in the $0.8-5.5$~$\mu$m range. Among the
different modes available in this instrument, we used the $0.8-2.4$~$\mu$m cross-dispersed mode (SXD),
which yields moderate spectral resolution ($R \sim 2000$) \citep{2003PASP..115..362R}. \\

In Table~\ref{tab:obs_17544} we present the set of observations of 
IGR~J17544-2619. We used an observation from SpeX taken on August~8,~2004. In 
the ESO archive there are 14 FEROS observations of IGR~J17544-2619 taken on four 
different dates during September~2005. There are not IUE available public 
observations of IGR~J17544-2619. \\

In Table~\ref{tab:obs_vela} we present the set of observations of Vela~X-1. In 
the ESO archive there are six consecutive FEROS observations of 700s taken on 
April~22,~2006. For the IUE data, we used the high dispersion and large aperture 
observations using the short-wavelength spectrograph (1150-2000~\AA) and the 
prime camera (SWP). There are 49 observations in the public database of the IUE 
following these criteria. \\

For each instrument, we averaged over all the available observations taking into account the exposure time
in order to improve the signal-to-noise ratio. We did not take the 
variability of the UV spectral lines depending on the orbital phase into account, 
that has been reported for Vela~X-1 \citep{1985ApJ...288..284S}. 
The variability consists on the presence of an extra absorption
component in several spectral lines, specially ones belonging to \AlIII ~and \FeIII ~, 
mainly at phases $\phi > 0.5$. This
variability must be taken into account to interpret the full picture of the stellar wind of Vela~X-1.
However, in this work, we prioritized a signal-to-noise ratio as high as possible over fitting a number
of phase dependent spectra with significantly lower signal-to-noise. This permits us to estimate the
stellar parameters of Vela~X-1 more accurately, while not affecting any of the conclusions derived in this
work, as we have carefully examined. \\

\begin{table}
\begin{tabular}{P{0.1\textwidth} | P{0.06\textwidth} P{0.1\textwidth} P{0.06\textwidth} P{0.06\textwidth} }
\toprule
Instrument & Phase & Date & MJD & Exposure \\
 & & {\scriptsize (YYYY-MM-DD)} & & (s) \\
\midrule
SpeX & 0.65 & 2004-08-15 & 53232.29 & 60 \\
\midrule
 & 0.01 & 2005-09-30 & 53643.05 & 1470 \\
 & 0.01 & 2005-09-30 & 53643.03 & 1470 \\
 & 0.01 & 2005-09-30 & 53643.01 & 1470 \\
 & 0.02 & 2005-09-30 & 53643.07 & 1470 \\
 & 0.61 & 2005-09-28 & 53641.08 & 1470 \\
 & 0.61 & 2005-09-28 & 53641.06 & 1470 \\
FEROS & 0.61 & 2005-09-28 & 53641.04 & 1470 \\
 & 0.62 & 2005-09-28 & 53641.10 & 1470 \\
 & 0.74 & 2005-09-09 & 53622.01 & 1470 \\
 & 0.75 & 2005-09-09 & 53622.02 & 1470 \\
 & 0.76 & 2005-09-09 & 53622.10 & 1470 \\
 & 0.76 & 2005-09-09 & 53622.08 & 1470 \\
 & 0.97 & 2005-09-15 & 53628.06 & 1470 \\
 & 0.98 & 2005-09-15 & 53628.08 & 1470 \\
\bottomrule
\end{tabular}
\caption{\footnotesize \label{tab:obs_17544} Table of observations of 
IGR~J17544-2619. We used $T_{90}=T_0=T_{\phi=0}=55924.271$~(MJD) and orbital 
period $P_\text{orb}=4.9272$~d \citep{2014MNRAS.439.2175D}. } 
\end{table}

\begin{table}
\begin{tabular}{P{0.1\textwidth} | P{0.06\textwidth} P{0.1\textwidth} P{0.06\textwidth} P{0.06\textwidth} }
\toprule
Instrument & Phase & Date & MJD & Exposure \\
 & & {\scriptsize (YYYY-MM-DD)} & & (s) \\
\midrule
  & 0.05 & 1978-05-05 & 43633.62 & 9000 \\      
  & 0.07 & 1984-02-19 & 45749.32 & 8280 \\
  & 0.08 & 1985-05-03 & 46188.67 & 4500 \\
  & 0.09 & 1985-05-03 & 46188.75 & 4500 \\
  & 0.09 & 1985-05-03 & 46188.81 & 3300 \\
  & 0.10 & 1985-05-03 & 46188.86 & 1020 \\
  & 0.10 & 1985-05-03 & 46188.92 & 6000 \\
  & 0.10 & 1993-11-08 & 49299.55 & 8400 \\
  & 0.14 & 1978-12-07 & 43849.51 & 8400 \\
  & 0.17 & 1992-11-06 & 48932.56 & 10800 \\
  & 0.22 & 1993-11-09 & 49300.55 & 8100 \\
  & 0.28 & 1983-01-22 & 45356.80 & 10800 \\
  & 0.28 & 1992-11-07 & 48933.57 & 9600 \\
  & 0.29 & 1983-01-22 & 45356.91 & 4500 \\
  & 0.29 & 1984-02-21 & 45751.31 & 9000 \\
  & 0.33 & 1993-11-10 & 49301.55 & 9000 \\
  & 0.40 & 1984-02-22 & 45752.36 & 9000 \\
  & 0.40 & 1988-02-22 & 47213.55 & 8460 \\
  & 0.41 & 1992-11-08 & 48934.72 & 9900 \\
  & 0.45 & 1978-04-30 & 43628.21 & 10800 \\
  & 0.46 & 1982-12-19 & 45322.52 & 9000 \\
  & 0.46 & 1993-11-11 & 49302.71 & 8400 \\
  & 0.49 & 1985-05-07 & 46192.36 & 7200 \\
  & 0.50 & 1985-05-07 & 46192.47 & 7200 \\
  & 0.51 & 1985-05-07 & 46192.58 & 7200 \\
 \large{SWP} & 0.52 & 1988-02-23 & 47214.54 & 8460 \\
  & 0.52 & 1988-03-12 & 47232.54 & 7826 \\
  & 0.53 & 1978-12-20 & 43862.03 & 7800 \\
  & 0.53 & 1983-01-07 & 45341.09 & 10800 \\
  & 0.55 & 1993-11-03 & 49294.56 & 6000 \\
  & 0.60 & 1978-12-02 & 43844.71 & 5400 \\
  & 0.61 & 1983-01-16 & 45350.77 & 10800 \\
  & 0.66 & 1993-11-04 & 49295.55 & 8400 \\
  & 0.71 & 1978-12-03 & 43845.69 & 8400 \\
  & 0.73 & 1984-02-16 & 45746.31 & 9000 \\
  & 0.74 & 1983-01-09 & 45343.01 & 10800 \\
  & 0.75 & 1985-04-21 & 46176.77 & 7200 \\
  & 0.76 & 1985-04-21 & 46176.86 & 4500 \\
  & 0.77 & 1979-03-21 & 43953.77 & 9000 \\
  & 0.77 & 1985-04-21 & 46176.99 & 6900 \\
  & 0.79 & 1993-11-05 & 49296.71 & 7500 \\
  & 0.84 & 1984-02-17 & 45747.32 & 9000 \\
  & 0.85 & 1978-07-23 & 43712.49 & 7500 \\
  & 0.90 & 1993-11-06 & 49297.73 & 6600 \\
  & 0.97 & 1983-01-11 & 45345.10 & 10800 \\
  & 0.97 & 1983-01-20 & 45354.07 & 5400 \\
  & 0.97 & 1984-02-18 & 45748.49 & 7500 \\
  & 0.98 & 1983-01-20 & 45354.13 & 3300 \\
  & 0.99 & 1993-11-07 & 49298.54 & 9600 \\
 \midrule  
  & 0.68 & 2005-04-22 & 53482.05 & 700 \\
  & 0.68 & 2005-04-22 & 53482.06 & 700 \\
 \large{FEROS} & 0.68 & 2005-04-22 & 53482.07 & 700 \\
  & 0.68 & 2005-04-22 & 53482.07 & 700 \\
  & 0.68 & 2005-04-22 & 53482.09 & 700 \\
  & 0.68 & 2005-04-22 & 53482.10 & 700 \\
\bottomrule	
\end{tabular}
\caption{\footnotesize \label{tab:obs_vela} Table of observations of Vela~X-1. 
We used $T_{90}=T_0=T_{\phi=0}=52974.001$~(MJD) and orbital period 
$P_\text{orb}=8.964357$~d \citep{2008A&A...492..511K}. }  
\end{table}

\section{The PoWR code \label{sec:PoWR}}
PoWR computes models of hot stellar atmospheres 
assuming spherical symmetry and stationary outflow. The non-LTE population 
numbers are calculated using the equations of statistical equilibrium and 
radiative transfer in the co-moving frame. Since these equations are coupled, 
the solution is iteratively found. Once convergence is reached, the synthetic 
spectrum is calculated integrating along the emergent radiation rays. The main 
features of the code have been described by \cite{2002A&A...387..244G} and 
\cite{2003A&A...410..993H}.  \\
	
The basic input parameters in PoWR are the following: stellar temperature 
($T_{\star}$), luminosity ($L_{\star}$), mass-loss rate ($\dot{M}$), surface 
gravity ($g_{\star}$) and chemical abundances. The chemical elements taken into 
account are detailed in Table~\ref{tab:chemic}. The stellar radius ($R_{\star}$) 
follows from $T_{\star}$ and $L_{\star}$ using the Stefan-Boltzmann law: 
$L_{\star} = 4\pi \sigma T_{\star}^4 R_{\star}^2$, where $\sigma$ is the 
Stefan-Boltzmann constant. We note that, in PoWR, $R_{\star}$ refers to the 
layer where the Rosseland continuum optical depth $\tau_\textrm{max}=20$, and not to 
the definition of stellar radius (or photospheric radius), where 
$\tau_\textrm{Ross}=2/3$. Nevertheless, we will give the stellar parameters in the next 
sections referring to both $\tau_\textrm{max}=20$ and the 
$\tau_\textrm{max}=2/3$, in order to avoid any confusion (e.g., we will use $R_{\star}$ 
for the radius at $\tau_\textrm{max}=20$ and $R_{2/3}$ for the radius at 
$\tau_\textrm{max}=2/3$). The surface gravity $g_{\star}$ and $R_{\star}$ imply the 
stellar mass ($M_{\star}$) via $g_{\star}= G M_{\star} R_{\star}^{-2}$. Instead 
of $g_{\star}$, one may specify the effective surface gravity $g_\text{eff}$, 
which accurately accounts for the outward force exerted by the radiation field, 
as thoroughly described by \cite{2015A&A...577A..13S}. \\

The density stratification in the stellar atmosphere, $\rho (r)$, is calculated 
from the continuity equation $\dot{M}= 4\pi r^2 \varv (r) \rho (r)$, given 
$\dot{M}$ and the radial velocity stratification $\varv (r)$. For $\varv (r)$, 
PoWR distinguish between two different regimes: the quasi-hydrostatic domain and 
the wind domain. A detailed description of the quasi-hydrostatic domain can be 
found in \cite{2015A&A...577A..13S}. In the wind domain, the $\beta$-law is 
adopted \citep{1975ApJ...195..157C}: 
\begin{equation}
\varv (r) = \varv_{\infty} \left( 1 - 
\frac{r_0}{r} \right)^{\beta}
\end{equation}
where $\varv_{\infty}$ is the terminal 
velocity of the wind, $r_0 \approx R_{\star}$ (depending on the precise location of the 
connection point) and $\beta$ is an input parameter typically ranging between 
$\beta = 0.6-2.0$ \citep{2008A&ARv..16..209P}. The connection point is chosen in 
order to ensure a smooth transition between the two domains. The temperature 
stratification is calculated from the condition of radiative equilibrium 
\citep{2003A&A...410..993H}. \\

The code also permits to account for density inhomogeneities and additional 
X-rays from a  spherically-symmetric, shock heated plasma. Density 
inhomogeneities are described in PoWR by means of an optional radial-dependent
input parameter: the density contrast $D(r)= \rho_{cl} / \bar{\rho}$, where 
$\rho_{cl}$ is the density of the clumped medium and $\bar{\rho}$ is the average 
density. The inter-clump medium is assumed to be empty. During the analysis, 
$D(r)$ is assumed to grow from $D(r_\text{sonic}) = 1$ (smooth plasma) to a 
maximum value $D$, which is reached at the layer where the stellar wind velocity 
is $\text{f}_\text{max} \times \varv_{\infty}$. $D$ is a free parameters derived in the analysis.
$\text{f}_\text{max}$ has a modest influence on the spectra. We assumed 
$\text{f}_\text{max} \sim 0.6$ on the basis of this moderate effect.
The X-rays are described using three parameters: the X-ray temperature 
$T_\text{X}$, the filling factor $X_\text{F}$ (i.e.\ the ratio between shocked 
to unshocked plasma), and the onset radius $R_{X}$, as described in 
\citet{1992A&A...266..402B}. In this work, 
we assumed $T_\text{X}=10^7$~K, $R_{X}=1.2\,R_{\star}$ and $X_\text{F}=0.05$. 
The main influence of X-rays in the model is via 
Auger ionization, which is responsible for the appearance of resonance lines belonging to 
high ions such as N{\sc v} and O{\sc vi} in the spectra of O~stars 
\citep{1979ApJ...229..304C, 2009MNRAS.394.2065K, 2011MNRAS.416.1456O}. 
Any changes in these parameters barely affect the
spectrum, as long as they they produce a similar X-ray luminosity. \\

During the iterative calculation of the population numbers, the spectral lines 
are taken to be Gaussian with a constant Doppler width of $\varv_\text{Dop} = 
40\,$km/s; the effect of $\varv_\text{Dop}$ on the spectrum is negligible for 
most lines \citep[see discussion by][]{2015ApJ...809..135S}. During the formal 
integration, the line profiles include natural broadening, pressure broadening, 
and Doppler broadening. The Doppler width is decomposed per element to a depth 
dependent thermal motion and a microturbulent velocity $\xi(r)$. The 
photospheric microturbulence, $\xi_\text{ph}$, is derived in the analysis, and 
beyond the photosphere we assumed that it grows from $\xi = \xi_\text{ph}$ to $\xi = 
100$~km/s at the layer where the stellar wind velocity is $500$~km/s. Rotational 
broadening is simulated via convolution with rotational profiles with a 
width corresponding to the projected rotational velocity $\varv_\text{rot}\,\sin i$ 
(denoted by $\varv_\text{rot}$ hereafter for simplicity), except for important wind
lines, for which the convolution is no longer 
valid \cite[see e.g.][]{2012MNRAS.426.1043H}, and where an explicit 
angle-integration would be required \citep[as described by][]{2014A&A...562A.118S}. The 
so-called macroturbulence $\varv_\text{mac}$ is accounted for by convolving the 
spectra with so-called Radial-Tangential profiles \citep{1975ApJ...202..148G, 
2007A&A...468.1063S}.

\section{The fitting procedure \label{sec:fit}}
We used the PoWR code to calculate synthetic spectra and a Spectral Energy 
Distribution~(SED) which best match the observations. The large number of free 
parameters, together with the long computation time for each model, do not 
permit the construction of a grid of models that covers the full parameter 
space. Instead, we attempted to identify the best-fitting model by visual 
inspection and systematic variation of the parameters. As an initial step, we 
calculate models using typical parameters of late~O / early~B stars. We then use 
specific spectral lines for each parameter as a guideline for the fit. 
Generally, the effective gravity $g_\text{eff}$ is derived from the 
pressure-broadened wings of the Balmer lines and He\,{\sc ii} lines. The 
temperature $T_\star$ is derived based on line ratios belonging to different ions of the 
same element. The mass-loss rate $\dot{M}$, $\varv_\infty$ and $D$ are derived 
from "wind-lines", with $D$ adjusted so that a simultaneous fit is obtained for 
both resonance lines (which scale as $\rho$) and recombination lines such as 
H$\alpha$ (which scale as $\rho^2$). The luminosity $L_\star$ and the reddening $E_{B 
- V}$ are derived by fitting the SED to photometry and flux-calibrated spectra. 
We apply the reddening law by \cite{1999PASP..111...63F}. Abundances are estimated from the overall
strengths of the spectral lines. The photospheric microturbulence $\xi_\text{ph}$ is found from the
strength and shape of helium lines. Finally, the parameters $\beta$, $\varv_\text{rot}$ and
$\varv_\text{mac}$ are adopted on the basis of the shape and depth of the spectral lines, together with
previous estimations found in the literature, when available. Upon adjusting the model, 
the whole spectral domain was examined
to iteratively improve the fit. Overall, we managed to find models which satisfactorily reproduce the
observed spectra and SEDs of the donors of the two systems analysed here. \\

We show the complete fits in Appendix~\ref{app:spectra}. The details about the fitting procedure for the
two objects are given in the following subsections. The obtained parameters are summarised in
Table~\ref{tab:params_fit} and the chemical abundances in Table~\ref{tab:chemic}. The parameters
that do not include an error estimation in the tables are adopted following the above mentioned criteria.\\ 

Even though the optical companion in Vela~X-1 is usually known as HD~77581, for 
the sake of simplicity we will refer to the donors with the name that is used 
for the X-rays sources, namely, IGR~J17544-2619 and Vela~X-1. Depending on the 
context, the reader should easily recognize whether it is the donor or the X-ray 
source which is being referred to. \\

\begin{table}
\begin{tabular}{P{0.14\textwidth} | P{0.14\textwidth} | P{0.14\textwidth} }
\toprule
Parameters & J17544-2619 & Vela~X-1 \\
\midrule
$\text{log}\,(L_\star/L_{\odot})$  & $  5.4\pm 0.1$ & $  5.5\pm 0.1$ \\
$M_{\star}/M_{\odot}$  & $ 25.9 \pm 2.0^b$ & $ 21.5\pm 4.0$ \\
$R_{\star} / R_{\odot}$  & $ 20_{-3}^{+4}$ & $ 28.4^a$ \\
$R_{2/3} / R_{\star}$  & $1.04$ & $ 1.09$ \\
$T_{\star}$ (kK)  & $ 29.0\pm 1.0$ & $ 25.5\pm 1.0$ \\
$T_{2/3}$ (kK)  & $ 28.5\pm 1.0$ & $ 24.4\pm 1.0$ \\
$\text{log}\,(g_{\star}\,\text{(cgs)})$  & $  3.25\pm 0.20$ & $  2.86\pm 0.10$ \\
$\text{log}\,(g_\text{eff}\,\text{(cgs)})$  & $  2.80\pm 0.20$ & $  2.35\pm 0.10$ \\
$\text{log}\,(g_\text{eff \, 2/3}\,\text{(cgs)})$  & $  2.77\pm 0.20$ & $  2.27\pm 0.10$ \\
$\varv_{\infty}$ (km/s) & $  1500\pm 200$ & $   700_{-100}^{+200}$ \\
$\varv_\text{esc}$ (km/s) & $   618\pm 75$ & $   436\pm 65$ \\
$\text{log}\,(\dot{M}/(M_{\odot}/yr))$ & $ -5.8\pm 0.2$ & $ -6.2\pm 0.2$ \\
$D$  & $  4 $ & $ 11 \pm 5$ \\
$\xi_\text{ph}$ (km/s) & $ 25 \pm 10$ & $  30 \pm 10$ \\
$\beta$ & $0.8$ & $1.0$ \\ 
$\varv_\text{mac}$ (km/s) & $60$ & $80$ \\
$\varv_\text{rot}$ (km/s) & $160$ & $56^c$ \\
$E_\text{B-V}$  & $2.14 \pm 0.10$  & $0.77 \pm 0.05$  \\
$R_\text{V}$ & $2.9$ & $3.1$ \\
$d$~(kpc)  & $3.0 \pm 0.2$  & $2.0 \pm 0.2$ \\
\bottomrule
\end{tabular}
\caption{\footnotesize\label{tab:params_fit} Stellar parameters obtained from 
the best fit.  (a)~\cite{1984ARA&A..22..537J}. (b)~\cite{2006A&A...455..653P}. (c)~\cite{2010MNRAS.404.1306F} } 
\end{table}

\begin{table}
\begin{tabular}{P{0.07\textwidth} | P{0.12\textwidth} P{0.035\textwidth} | P{0.12\textwidth}  P{0.035\textwidth} }
\toprule
 & \multicolumn{2}{c|}{IGR~J17544-2619} & \multicolumn{2}{c}{Vela~X-1}\\
\cmidrule(r){2-3} \cmidrule(r){4-5}
 Quemical Element  & Mass Fraction & Rel. Ab. & Mass Fraction & Rel. Ab.\\
\midrule
$\text{H}$  & $  (6.2 \pm 0.5)E-01$ & $ 0.85$ & $  (6.5 \pm 0.5)E-01$ & $ 0.89$ \\
$\text{He}$  & $  (3.7 \pm 0.5)E-01$ & $ 1.47$ & $  (3.4 \pm 0.5)E-01$ & $ 1.35$ \\
$\text{C}$  & $  (5.0 \pm 3.0)E-04$ & $ 0.17$ & $  (5.0 \pm 3.0)E-04$ & $ 0.17$ \\
$\text{N}$  & $  (2.2 \pm 0.6)E-03$ & $ 2.58$ & $  (1.8 \pm 0.6)E-03$ & $ 2.11$ \\
$\text{O}$  & $  (6.0 \pm 2.0)E-03$ & $ 0.76$ & $  (7.0 \pm 0.2)E-03$ & $ 0.88$ \\
$\text{Si}$  & $ (7.3 \pm 2.0)E-04$ & $ 1.00$ & $  (5.5 \pm 2.0)E-04$ & $ 0.75$ \\
$\text{S}$  & $  5.0E-04$ & $ 1.00$ & $  5.0E-04$ & $ 1.00$ \\
$\text{P}$  & $  6.4E-06$ & $ 1.00$ & $  6.4E-06$ & $ 1.00$ \\
$\text{Al}$  & $  5.8E-05$ & $ 1.00$ & $  7.0E-05$ & $ 1.00$ \\
$\text{Mg}$  & $  7.0E-04$ & $ 1.00$ & $  7.0E-04$ & $ 1.00$ \\
$\text{Fe}^a$  & $  1.4E-03$ & $ 1.00$ & $  1.4E-03$ & $ 1.00$ \\
\bottomrule
\end{tabular}
\caption{\footnotesize \label{tab:chemic} Chemical abundances derived from the 
best fit, in mass fraction and relative to solar abundances from \cite{2009ARA&A..47..481A}. 
(a)~The notation of Fe actually stands for a generic atom including iron group elements Sc, Ti, V, Cr, Mn, 
Co and Ni. For more details see \cite{2002A&A...387..244G}.  } 
\end{table}

\subsection{IGR~J17544-2619}
IGR~J17544-2619 was first detected on September 2003 with the IBIS/ISGRI detector on board
\textit{INTEGRAL} \citep{2003ATel..190....1S}. It is located in the direction of the galactic center, at
galactic coordinates $l=3.24^{\circ}$, $b=-0.34^{\circ}$. The orbital period is $\sim$4.9d
\citep{2009MNRAS.399L.113C}. According to \textit{Chandra} observations, the compact object is a neutron
star \citep{2005A&A...441L...1I}. \cite{2006A&A...455..653P} used optical and NIR observations in order to
classify the optical companion as a O9Ib. 
\textit{Chandra} and \textit{Swift} observations showed that
the system exhibits a high dynamic range in its X-ray variability, changing the X-ray flux by 5
orders of magnitude \citep{2005A&A...441L...1I,2015A&A...576L...4R}.  \\

Nowadays, the spin period $P_\text{spin}$ of the hypothetical neutron star in IGR~J17544-2619 is 
a matter of debate, given the results arising from observations taken at different times, different luminosities
and different instruments. \cite{2012A&A...539A..21D} analysed RXTE data of the source at intermediate X-ray
luminosity ($\sim 10^{33-34}$~erg/s), and reported the detection of an X-ray pulsation with $P_\text{spin}=71.49$s
at a statistical significance of $4.37\sigma$. 
\cite{2015A&A...576L...4R} inspected \textit{Swift} observations of the source experiencing an extraordinarily bright outburst
(peak luminosity $\sim 10^{38}$~erg/s), and reported the detection of X-ray pulsations with $P_\text{spin}=11.60$s
at a statistical significance of about $4\sigma$ too. However, these results contrast with the analyses of \textit{XMM-Newton} and \textit{NuSTAR} observations performed by \cite{2014MNRAS.439.2175D} and \cite{2015MNRAS.447.2274B} respectively. These authors do not find any evidence of pulsations on time scales of 1-2000s. \\ 

\begin{figure}
\centering
\epsfig{file=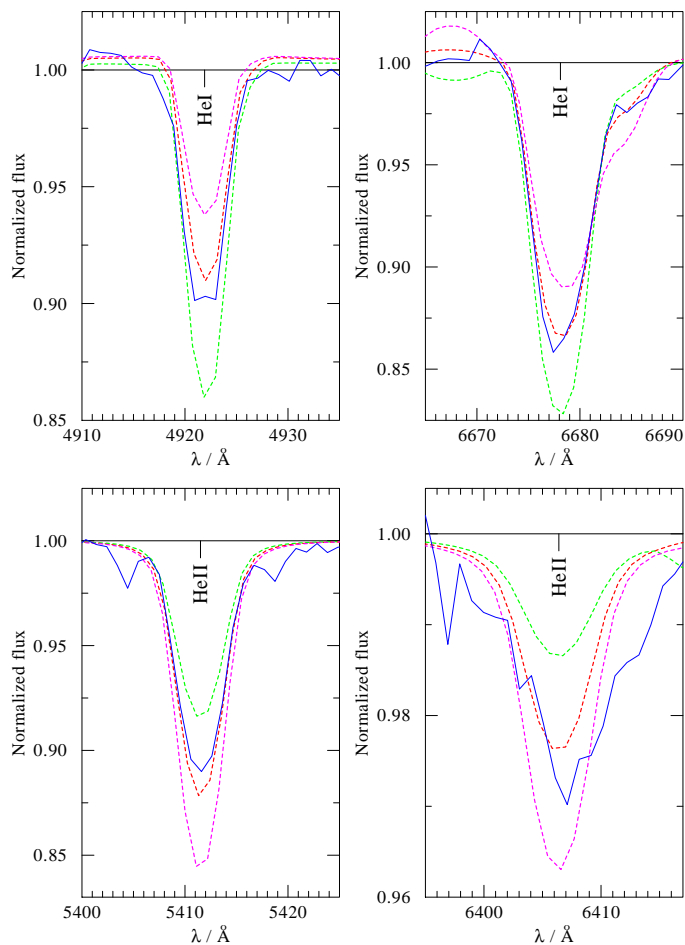, width=0.48\textwidth}
\caption{\footnotesize Example of four helium lines in IGR~J17544-2619, used for 
the estimation of T$_{\star}$. We show the observation (solid blue line), the 
best-fit model (red dashed line), a model with lower temperature of
$T_{\star}=28$~kK (green dashed line), and a model with higher temperature of
$T_{\star}=30$~kK (pink dashed line). }
\label{fig:Teff_17544}
\end{figure}

We have adjusted T$_{\star}$ of IGR~J17544-2619 using different ions, mainly 
\HeI-\HeII ~and \SiIII-\SiIV. In Fig.~\ref{fig:Teff_17544} we show an example 
of four helium lines of which the best-fit model provides a good description. 
Higher (lower) temperatures yield more (less) absorption than observed in the 
\HeII ~lines. We have used other lines of helium, silicon, nitrogen and oxygen. 
The vast majority of them are well described by the best-fit model, within the 
errors. The obtained effective temperature is compatible 
with the donor's spectral class O9~Ib \citep{2005A&A...436.1049M}. \\

\begin{figure}
\centering
\epsfig{file=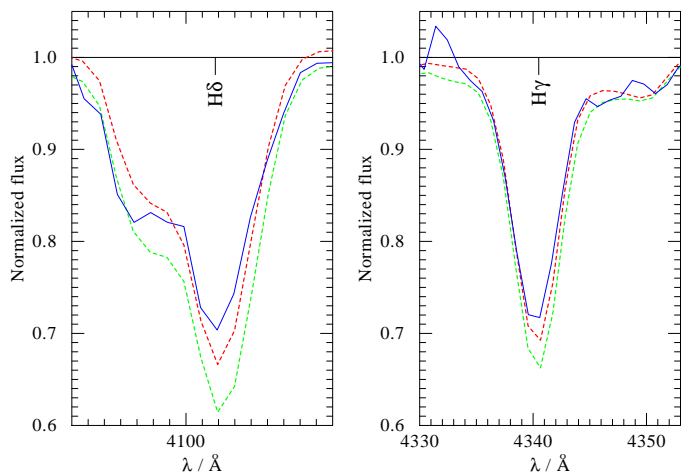, width=0.48\textwidth}
\caption{\footnotesize H$\gamma$ and H$\delta$ in IGR~J17544-2619, used for the 
surface gravity estimation. We show the observation (solid blue line), the 
best-fit model (red dashed line) and a model with larger effective gravity of $\text{log}\,(g_\text{eff})=3.0$ in cgs units (green
dashed line). }
\label{fig:geff_17544}
\end{figure}

The effective gravity g$_\text{eff}$ was found using the hydrogen 
Balmer lines H$\gamma$ and H$\delta$. We did not use H$\beta$ and H$\alpha$ 
because these lines are notably affected by the stellar wind. 
Figure~\ref{fig:geff_17544} shows a comparison of the observations with the 
best-fitting model for these two Balmer lines. We show that the observations are 
compatible with a relatively wide range of values, as also reflected in the 
errors given in Table~\ref{tab:params_fit}. \\

The distance to IGR~J17544-2619 is not well known, with an estimate of 2-4~kpc 
\cite{2006A&A...455..653P}, based on the extinction and the calibration of the 
absolute magnitude for O9Ib stars. In this work we improve this estimation. 
As a first step, we fitted the SED to photometry from the 2MASS 
catalogue \citep{2003yCat.2246....0C}, \cite{2012yCat.1322....0Z} and 
\cite{Rahoui2008} assuming the distance to be 3~kpc. Then,
we derived initial values for the luminosity of the donor and 
the reddening to the system. \\

\begin{figure}
\epsfig{file=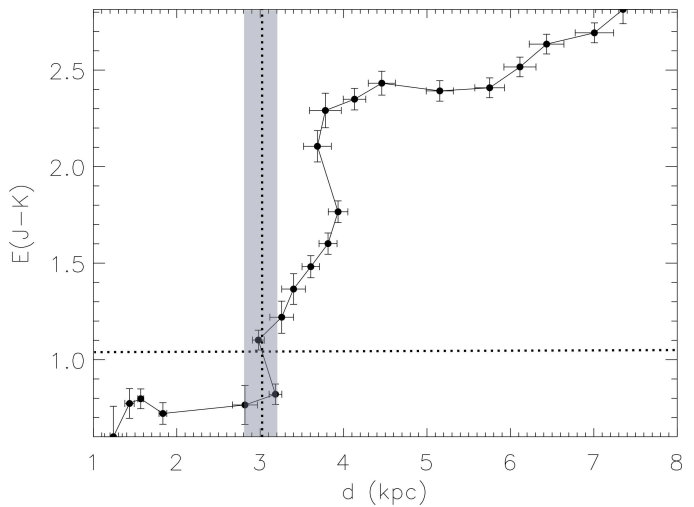, angle=180, width=0.48\textwidth}
\caption{\footnotesize Extinction curve in the galactic direction of IGR~J17544-2619. The 
shaded area reflects the error in the distance estimation from the errors of 
estimation of the extinction and the errors in the calculation of the extinction 
curve. }
\label{fig:extin}
\end{figure}

\begin{figure*}
\epsfig{file=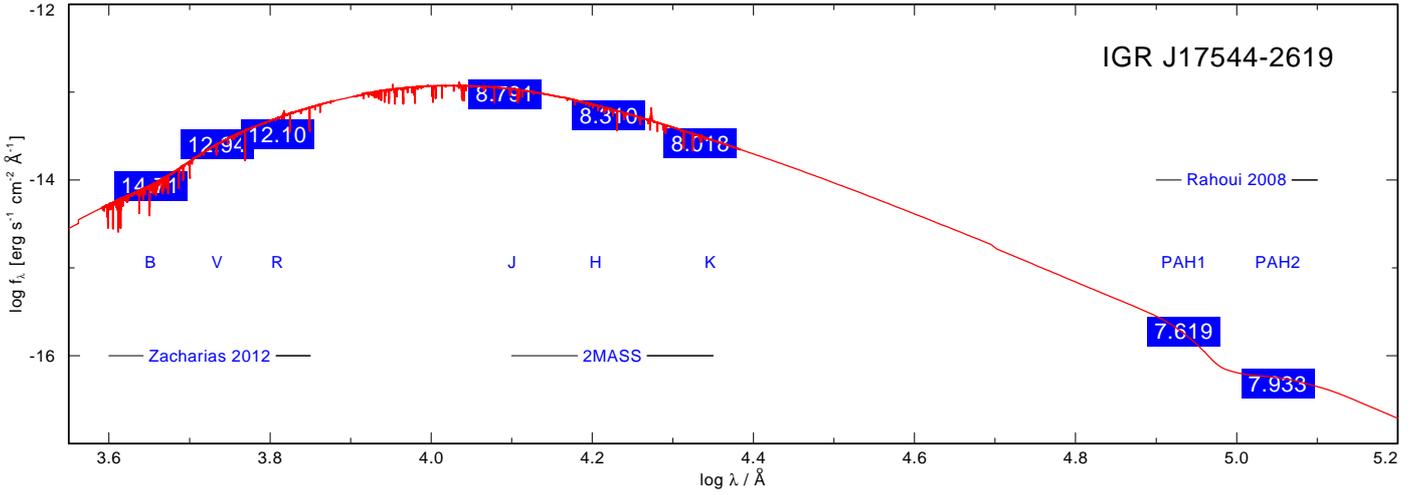, angle=-90, width=\textwidth}
\caption{\footnotesize Fit of the SED of IGR~J17544-2619. In red we plot the best-fit model. We indicate
the photometry values for each band in blue. The employed references are cited. The values of extinction,
distance and luminosity are shown in Table~\ref{tab:params_fit}. }
\label{fig:SED_17544}
\end{figure*}

As a second step, in order to provide more constraints on the distance, we employed a method
based on the well constrained luminosity of Red Clump Giant stars (RCG). These stars can be isolated
in a NIR colour-magnitude diagram and permit the estimation of the interstellar extinction
along the line of sight \citep{2002A&A...394..883L}. Due to their narrow luminosity function,
the apparent magnitude of RCGs provides an estimation of the distance. Then, given a certain line of sight, 
a diagram of the extinction versus the distance can be derived \citep[for more details see][]{2014ApJ...782...86G}.
For IGR~J17544-2619 we employed the derived $E_{J-K}$ from the SED fit to obtain an estimate of the distance. 
We note that this method is only applicable to
stars in the direction of the galactic center like IGR~J17544-2619, where the medium is more homogeneous and the
density of RCGs is higher. Using this method, we obtain a distance of $3.0 \pm 0.2$~kpc (Fig.~\ref{fig:extin}). 
Revised values for the luminosity and reddening are then derived. The final results of the 
SED fit are shown in Fig.~\ref{fig:SED_17544}. \\


From the luminosity and temperature we derive $R_{\star}$, which provides an upper limit 
to the eccentricity of the system. For the lower limit $R_\star = 17\,R_\odot$, 
we find $e<0.25$. For higher eccentricities, periodic Roche-lobe overflow is 
expected from the orbital solution of the system \citep{2009MNRAS.399L.113C}, 
at odds with the X-ray behavior of the source. Given the radius of the 
source and the derived surface gravity, we find $M_{\star}=25.9\,M_\odot$. This value matches very well with the estimation of $M_{\star}=25-28\,M_\odot$ done by \cite{2006A&A...455..653P} based on the mass calibration with its spectral type. \\

\begin{figure}
\centering
\epsfig{file=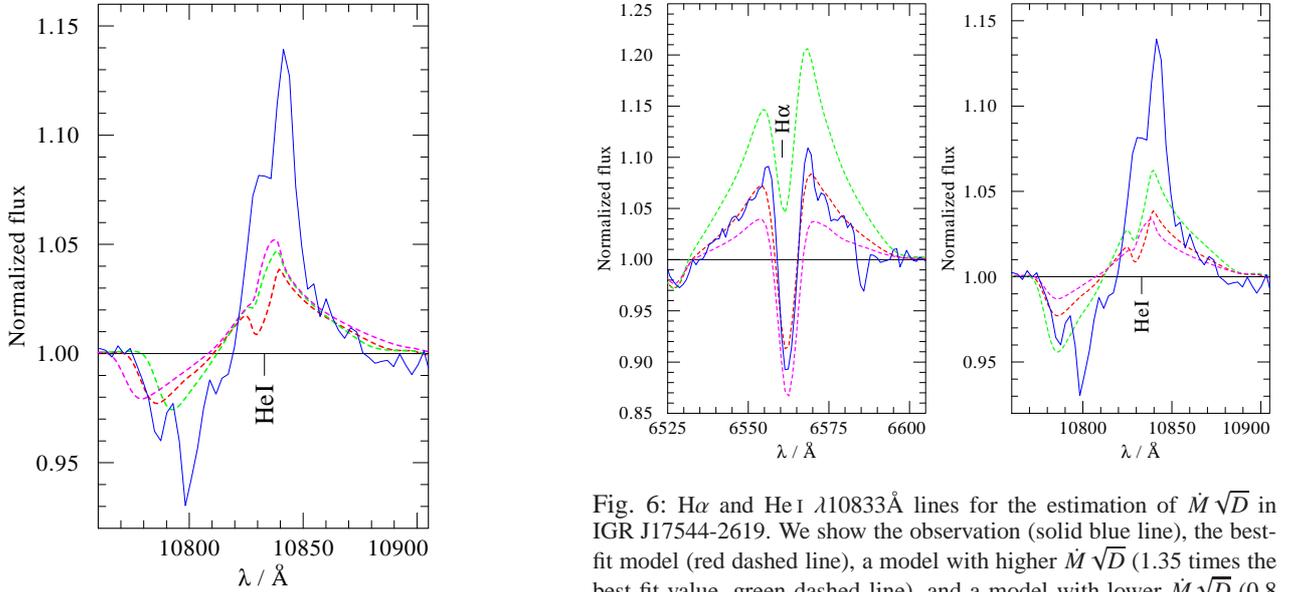, width=0.3\textwidth}
\caption{\footnotesize \HeI ~at $\lambda \, 10833\,\AA$, used for 
the estimation of $\varv_{\infty}$ in IGR~J17544-2619 fitting the blue wing of the P-Cygni 
profile. We show the observation 
(solid blue line), the best-fit model (red dashed line), a model with 
$\varv_{\infty}=1300$~km/s (green dashed line), and a model with 
$\varv_{\infty}=1700$~km/s (pink dashed line).   }
\label{fig:Vinf_17544}
\end{figure}

The terminal velocity of the stellar wind $\varv_{\infty}$ was derived using the 
P-Cygni profile of \HeI ~$\lambda 10833\,\AA$ (see 
Fig.~\ref{fig:Vinf_17544}). The blue wing in \HeI ~$\lambda 
10833\,\AA$ is a very good indicator due to its strong sensitivity to 
$\varv_{\infty}$. It is reasonably well fitted when assuming $\varv_{\infty} 
\simeq 1500$~km/s. Unfortunately, the emission exhibited by this line is not
well reproduced by the best-fit model, as explained below.  \\

\begin{figure}
\centering
\epsfig{file=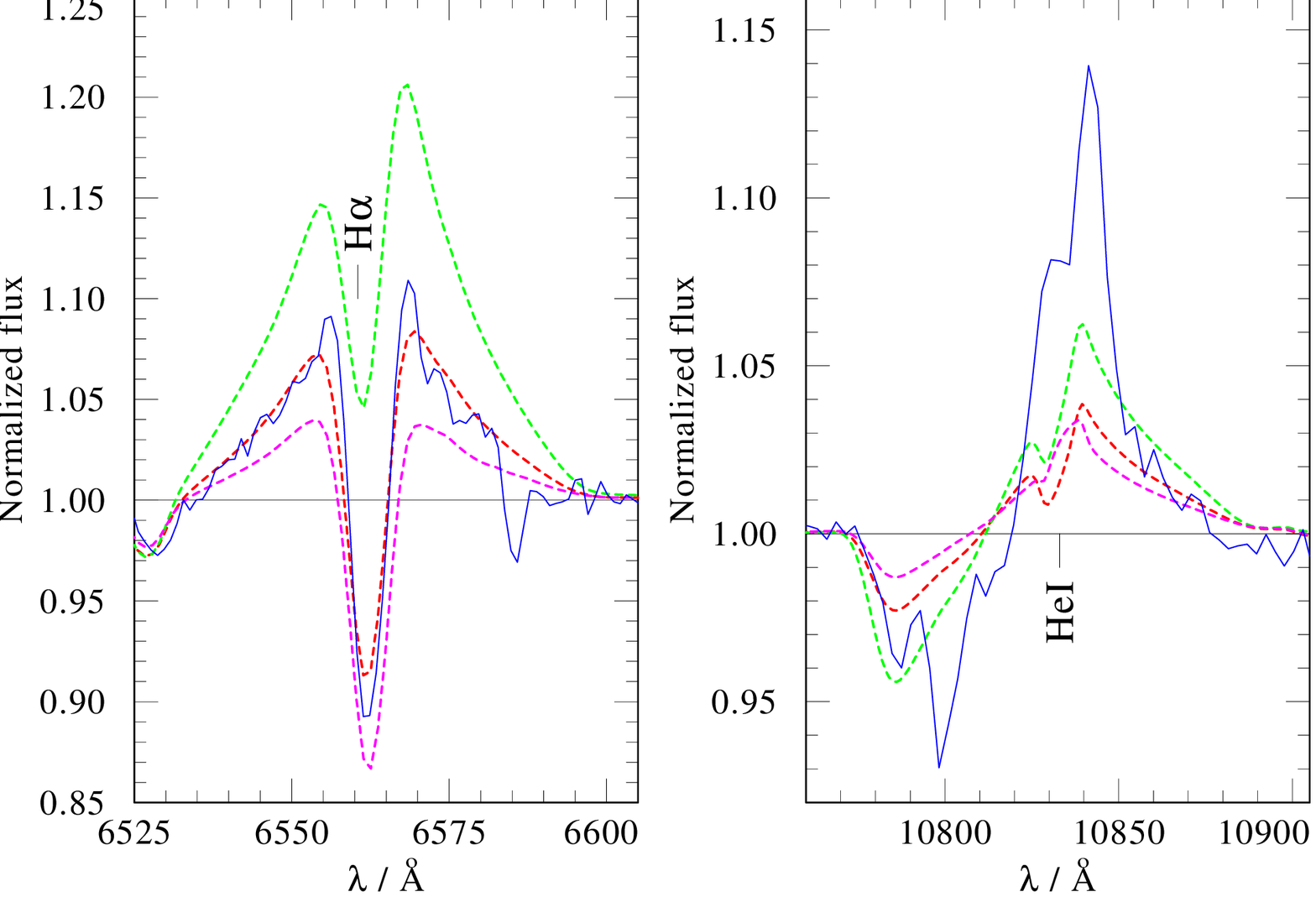, width=0.48\textwidth}
\caption{\footnotesize H$\alpha$ and \HeI ~$\lambda 10833\AA$
lines for the estimation of $\dot{M} \sqrt{D}$ in IGR~J17544-2619. We show the 
observation (solid blue line), the best-fit model (red dashed line), a model 
with higher $\dot{M} \sqrt{D}$ ($1.35$ times the best-fit value, green dashed line), and a model
with lower $\dot{M} \sqrt{D}$ ($0.8$ times the best-fit value, pink dashed line). }
\label{fig:Rt_17544}
\end{figure}

\begin{figure}
\centering
\epsfig{file=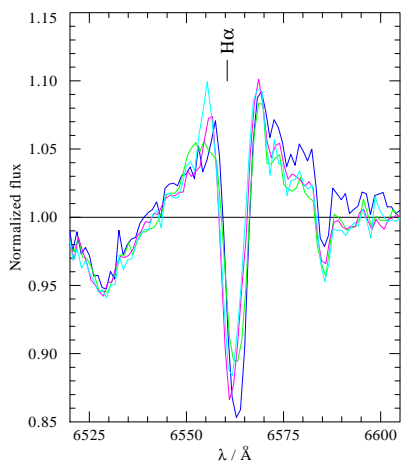, width=0.28\textwidth}
\caption{\footnotesize H$\alpha$ in IGR~J17544-2619 at different orbital phases: 
$\phi \simeq 0.01,0.61,0.75,0.97$ (blue, green, pink and turquoise solid lines 
respectively). }
\label{fig:Ha_var_17544}
\end{figure}

The $\dot{M}$ and $D$ were simultaneously adjusted by means of H$\alpha$ 
and the P-Cygni profile of HeI~$\lambda 10833$~\AA. Provided that 
the strength of emission in these recombination spectral lines varies with 
$\dot{M} \sqrt{D}$ \citep{2002A&A...387..244G}, we cannot estimate 
$\dot{M}$ and $D$ independently using these lines. As it is shown in Fig.~\ref{fig:Rt_17544}, we 
were not able to fit all the lines at the same time. The best-fit model provides 
an acceptable description of H$\alpha$, but yields insufficient emission for 
HeI~$\lambda 10833$~\AA. We choose the best description of H$\alpha$ as the 
best-fit because it provides a better fit to the overall spectrum. We note that the optical 
and infrared spectra were not taken at the same time, and therefore any kind of 
variability in the lines might produce a disagreement. However, H$\alpha$ does 
not show such a large variability within the observations we have analysed (see 
Fig.~\ref{fig:Ha_var_17544}).  \\

\begin{figure}
\centering
\epsfig{file=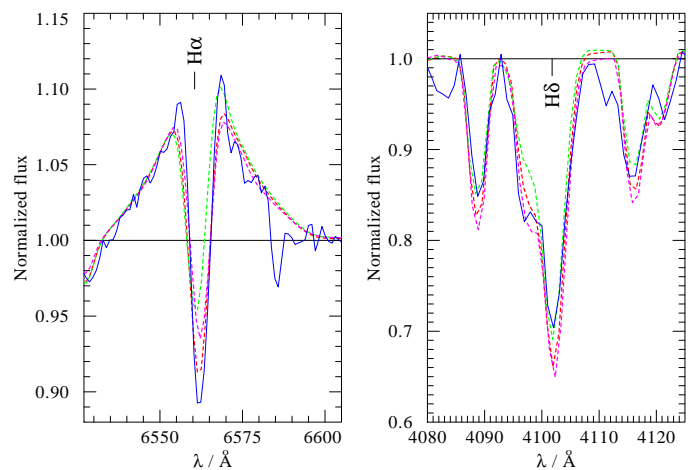, width=0.48\textwidth}
\caption{\footnotesize H$\alpha$ and H$\delta$ 
~for the estimation of $\dot{M}$ in IGR~J17544-2619. We show the observation 
(solid blue line), the best-fit model ($\dot{M}=10^{-5.8} \, M_{\odot}/yr$, 
$D=4$, red dashed line), a model with higher $\dot{M}$ ($\dot{M}=10^{-5.5} \, 
M_{\odot}/yr$, $D=1$, green dashed line), and a model with lower $\dot{M}$ 
($\dot{M}=10^{-5.9} \, M_{\odot}/yr$, $D=8$, pink dashed line). Different $\dot{M}$ 
values do not yield different H$\alpha$ wings as long as the product $\dot{M} \sqrt{D}$ remains 
constant. However, we observed that other important lines like H$\delta$ are sensitive these variations. }
\label{fig:Mdot_17544}
\end{figure}

Without available resonance scattering lines in the observations at hand, we cannot compare 
P-Cygni lines with recombination lines to deduce the clumping factor $D$. However, our calculations show
that changing $\dot{M}$ dramatically affects the absorption spectrum in a fashion which is not related to
the product $\dot{M} \sqrt{D}$. An example is shown in Fig.~\ref{fig:Mdot_17544}, where we 
show three models calculated with different values of $\dot{M}$ and $D$, but with a fixed product $\dot{M}
\sqrt{D}$. Evidently, while the emission exhibited by the wings of H-$\alpha$ (shown in
Fig.~\ref{fig:Mdot_17544}) is similar
in all models, the absorption lines are strongly affected in a non-trivial manner. 
The reason for this unexpected behaviour is that many of the strong lines in the spectrum (e.g.\ the
Balmer series) are formed significantly beyond the photosphere ($\tau_{\rm Ross} \approx 2/3$), where the
mass-loss rate already strongly affects the density stratification via the continuity equation.
Exploiting this effect, we find that $D \approx 4$ provides the best results for the overall spectrum.
However, we warn that further observations are needed to better constrain the clumping factor in this star.
Nevertheless, we note that our final conclusions do not strongly depend on this factor and the
implied mass-loss rate,
as will be discussed in Section\,\ref{sec:disc}. \\  

The chemical composition was estimated from unblended spectral lines for He, C, 
N, O and Si. The rest of the considered element abundances (see Table~\ref{tab:chemic})
were assumed solar following \cite{2009ARA&A..47..481A}.  The fit yielded moderate overabundance 
of He and N, together with underabundance of C and O. In all, there are 
indications of chemical evolution in the outer layers of the stellar atmosphere.  
\\ 

\begin{figure}
\centering
\epsfig{file=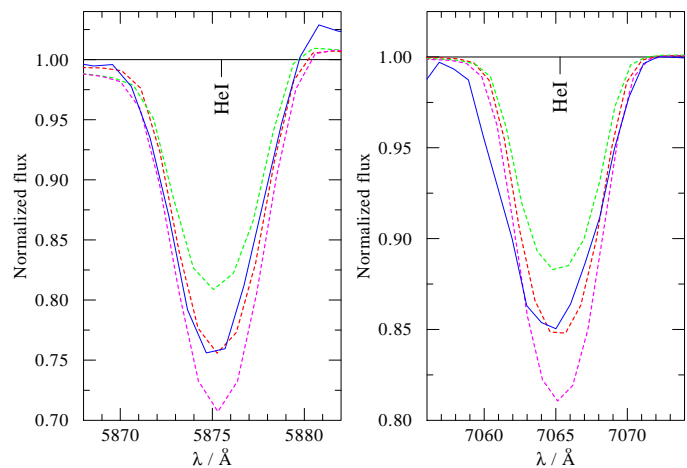, width=0.48\textwidth}
\caption{\footnotesize Example of two \HeI ~lines in IGR~J17544-2619, used in the 
$\xi_\text{ph}$ estimation. As usual, the observation is plotted in solid blue line, and 
the best-fit model in red dashed line. Models with $\xi_\text{ph}=15,35$~km/s are also 
presented (green and pink dashed lines respectively).  }
\label{fig:vmic_17544}
\end{figure}

The photospheric microturbulent velocity ($\xi_\text{ph}$) was adjusted using \HeI ~and \SiIV ~lines. 
A higher $\xi_\text{ph}$ induce stronger absorption in several spectral lines, as shown in 
Fig.~\ref{fig:vmic_17544}. \\

The $\varv_\text{rot}$ and $\varv_\text{mac}$ were roughly estimated using the width of the He 
lines. The derived projected rotational velocity is around 0.3~times the 
critical rotation velocity ($\varv_\text{crit} = \sqrt{G \, M_\star R^{-1}_\star}$). This high rotational velocity may favour the chemical mixing, in line with the abundances derived in the fit. \\

To summarise, our NLTE analysis of optical and near IR spectra of IGR~J17544-2619 
showed that the optical O9I-type companion in this source is not peculiar and has stellar and wind parameters that are similar to other stars of the same spectral type, e.g. $\delta$~Ori \citep{2015ApJ...809..135S}. 

\subsection{Vela~X-1 \label{sec:vela}}
Vela~X-1 is one of the most studied HMXBs, since it is a bright source discovered in the early ages of the
X-ray astronomy \citep{1967ApJ...150...57C}. It is located at galactic coordinates $l=263.06^{\circ}$,
$b=3.93^{\circ}$. The distance was estimated to be $1.9 \pm 0.2$~kpc by \citep{1985ApJ...288..284S}. The
system has a moderate eccentricity of $e=0.09$ \citep{1997ApJS..113..367B}, and orbital period
$P_\text{orb}=8.96$~days \citep{2008A&A...492..511K}. The compact object is a neutron star that pulsates
with $P_\text{spin}=283$s \citep{1976ApJ...206L..99M}. The optical companion HD~77581 (B0.5Iae) was
identified by \citet{1973ApJ...182L..77V}. \\

It is very likely that the wind of Vela~X-1 is disturbed by the X-ray source. 
The photoionization produced close to the photosphere due to the intense X-ray 
luminosity might hinder the acceleration of the wind and generate a structure 
known as photoionization wake \citep{Blondin1990, 2015A&A...579A.111K}. 
This structure appears in the UV spectra as an additional absorption component at phases larger 
than $\phi \sim 0.5$ \citep{1994A&A...289..846K}. In addition, 
the hard X-rays light curves of the source in 
near-to-eclipse phases show asymmetries between ingress and egress, that have been interpreted as
caused by the existence of this type of structure trailing the neutron star \citep{1996A&A...311..793F}. 
Moreover, a density enhancement in the line of sight during the 
second half of the orbit is also observed in the X-rays absorption, although the 
amount of absorbing material is highly variable from one orbit to another.  \\

We derived $T_\star$ following the same procedure that we used for 
IGR~J17544-2619. The obtained $T_{2/3}$ is similar to previous estimations: 
\cite{1985ApJ...288..284S} used the equivalent width (EW) of photospheric lines 
to estimate the effective temperature $T_{2/3}=25000$~K; 
\cite{2010MNRAS.404.1306F} used the TLUSTY code to estimate $T_{2/3}=26500$~K. \\

\begin{figure*}
\epsfig{file=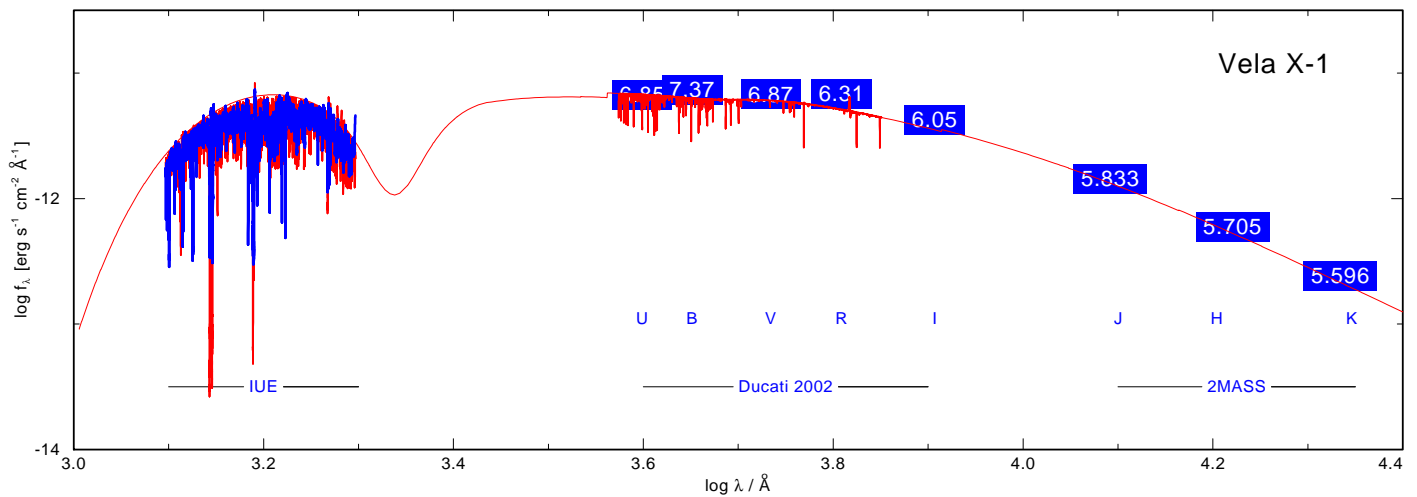, angle=-90, width=\textwidth}
\caption{\footnotesize Fit of the SED of Vela~X-1. In red we plot the best fit model with the spectral
lines in the domain where we have done the spectral analysis, and the continuum where we have available
photometry (marked in blue). We cite the references used for the photometry. Note that the true continuum
in the UV range do not correspond with the apparent continuum from the observation, due to the number of
spectral lines in this domain. The employed values of extinction, distance and luminosity are shown in
Table~\ref{tab:params_fit}.  }
\label{fig:SED_Vela}
\end{figure*}

For the fit of the SED, we used photometry from the 2MASS catalogue 
\citep{2003yCat.2246....0C} and the \textit{Stellar Photometry in Johnson's 
11-color system} \citep{2002yCat.2237....0D}, together with the IUE 
observations. We made an estimation of the reddening, distance and $R_\text{V} \equiv 
A(V)/E_{B-V}$ by means of the SED fit. Then, we used the estimation of the 
stellar radius $R_{2/3}=31\,R_{\odot}$ from \cite{1984ARA&A..22..537J}, and 
$T_{2/3}$ from the successive fits, in order to derive the luminosity (and the 
distance estimation) from the Stefan-Boltzmann law. Given that the obtained 
T$_{2/3}$ is very similar to previous estimations, the derived distance of $2.0 
\pm 0.2$ is almost equal to the value d=1.9~kpc given by 
\cite{1985ApJ...288..284S}. We show the results of the SED analysis in 
Fig.~\ref{fig:SED_Vela}. \\

The estimation of g$_\text{eff}$ was especially delicate in Vela~X-1 because of its 
very low g$_\text{eff}$. A higher value beyond the error given in Table~\ref{tab:params_fit} has a strong 
effect in the overall spectrum and hinders a satisfying fit. The 
derived value enables a good fit, and it is in agreement with previous 
estimations \citep{2010MNRAS.404.1306F}. \\

\begin{figure}
\centering
\epsfig{file=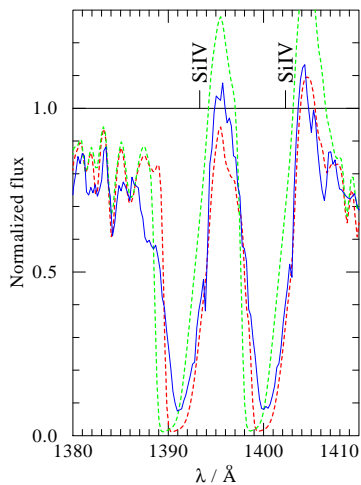, width=0.25\textwidth}
\caption{\footnotesize \SiIV ~lines used for the estimation of $\varv_{\infty}$ 
in Vela~X-1. It is showed the observation (blue solid line), the best-fit model 
(red dashed line) and a model with $\varv_{\infty}=900$~km/s (green dashed line). }
\label{fig:Vinf_Vela}
\end{figure}

We used UV resonance lines to find $\varv_{\infty}$. In 
Fig.~\ref{fig:Vinf_Vela} we show the \SiIV ~resonance lines $\lambda \, 1394,1403\,\AA$, where the
effect of $\varv_{\infty}$ is very clear. Models with higher terminal velocities induce 
a shift towards the blue part of these spectral lines. The best description of 
the observations is achieved for $\varv_{\infty}=700$~km/s. This value is in 
agreement with the estimation of \cite{2001A&A...375..498V}: 
$\varv_{\infty}=600$~km/s; and not too far from \cite{2006ApJ...651..421W}, who 
estimated $\varv_{\infty}=1100$~km/s using \textit{Chandra} X-rays observations. \\

\begin{figure}
\centering
\epsfig{file=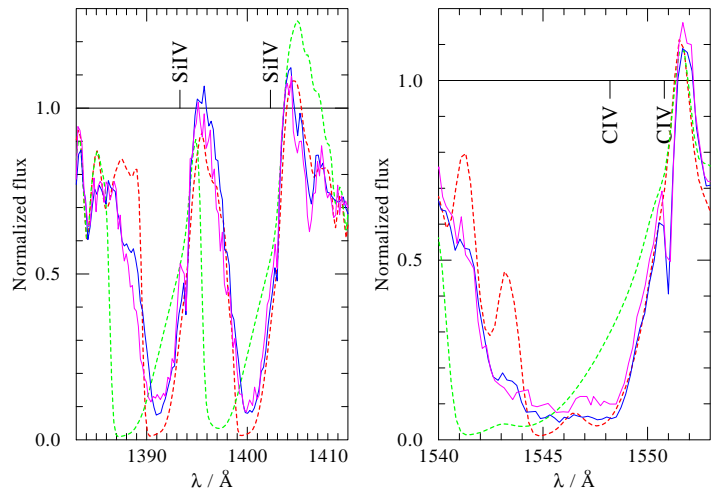, width=0.5\textwidth}
\caption{\footnotesize \SiIV ~and \CIV ~resonance lines. We plot the total averaged spectrum (blue solid line) and the averaged spectrum over orbital 
phases $\phi=0.9-0.1$ (pink solid line). We also plot the best-fit model 
($\varv_{\infty}=700$, red dashed line), and a model with 
$\varv_{\infty}=1400$~km/s (green dashed line).  }
\label{fig:Dupree}
\end{figure}

In contrast, it is in disagreement with the estimation of \cite{1980ApJ...238..969D}, namely
$\varv_{\infty}=1700$~km/s. These authors used a subset of the IUE observations used in this
work, and considered the UV resonance lines \SiIV ~and \CIV ~in the X-ray eclipse phases to make their
estimation. We have revisited our $\varv_{\infty}$ estimation using only observations taken at orbital
phases $\phi=0.9-0.1$, in order to be able to directly compare to \cite{1980ApJ...238..969D}. In
Fig.~\ref{fig:Dupree}, we show the \SiIV ~and \CIV ~lines, as observed in the total averaged spectrum
and the spectrum averaging over $\phi=0.9-0.1$. \CIV ~is almost the same in both cases. Then, the
disagreement in the estimates of $\varv_{\infty}$ does not come from orbital phase variations but from
the omission of the impact of the X-rays in the stellar wind by \cite{1980ApJ...238..969D}. 
As we can see in Fig.~\ref{fig:Dupree}, when
we introduce X-rays in the models we are able to reproduce \CIV ~without needing a high velocity, due 
to the significant enhancement of the population of \CIV ~in the wind. We note that the X-ray radiation we
are introducing in the models is an intrinsic radiation of the donor wind 
that is presumably produced in the shocks within 
the stellar wind itself \citep[e.g.][]{2009A&A...508..841K}. This radiation is
not coming from the neutron star, since the effects are also noticeable at eclipsing phases. The impact of
the X-rays coming from the neutron star is a different and complex issue, and it has been already studied
by other authors \citep{2006ApJ...651..421W}. Regarding the \SiIV ~resonance lines, in
Fig.~\ref{fig:Dupree} we show that high stellar wind velocities as derived by \cite{1980ApJ...238..969D}
do not fit, neither using the total averaged spectrum, neither using the eclipsing phases spectrum.   \\

\begin{figure}
\centering
\epsfig{file=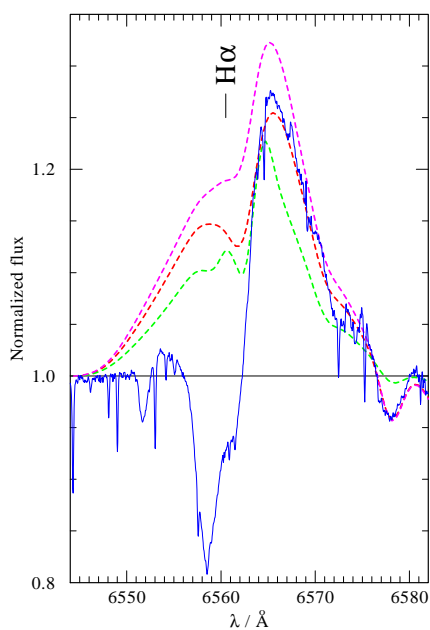, width=0.3\textwidth}
\caption{\footnotesize H$\alpha$ line for the estimation of $\dot{M} \sqrt{D}$ 
in Vela~X-1. We can see the observation (blue solid line), the best-fit model 
(red dashed line), a model with $0.8$ times the $\dot{M} \sqrt{D}$ value of the 
best fit (green dashed line), and a model with $1.2$ times the $\dot{M} \sqrt{D}$ value of the 
best fit (pink dashed line). }
\label{fig:Rt_Vela}
\end{figure}

The value $\dot{M} \sqrt{D}$ was estimated using H$\alpha$ (see 
Fig.~\ref{fig:Rt_Vela}). We did not find a good fit of the blue wing of the 
line, observed in absorption, but our model properly fits the emission in the 
red wing of the spectral line. Unfortunately, we do not have more optical 
observations covering further orbital phases in order to check whether H$\alpha$ is 
variable. Nevertheless, previous studies of similar sources demonstrate that 
this might be the case: \cite{2015arXiv150301087G} reported the variability of 
H$\alpha$ in the very similar B0Iaep optical companion in the SGXB system 
XTE~J1855-026. Moreover, the shape of H$\alpha$ in XTE~J1855-026 at $\phi=0$ 
\citep[see Fig.~5.12 in][]{2015arXiv150301087G}, when the neutron star is hidden 
behind the optical counterpart, is strongly reminiscent of the shape that our 
model reproduces in Fig.~\ref{fig:Rt_Vela}. Hence, the relative disagreement 
between our best-fit model and our observation of Vela~X-1 (taken at 
$\phi=0.68$), might be produced by some kind of interaction of the neutron star 
with the donor and/or the stellar wind, which is not possible to model using the 
assumption of spherical symmetry that PoWR employs. This disagreement might be related
to similar features observed in other strong lines, as further discussed in Sect.~\ref{sec:asym}. \\

\begin{figure}
\centering
\epsfig{file=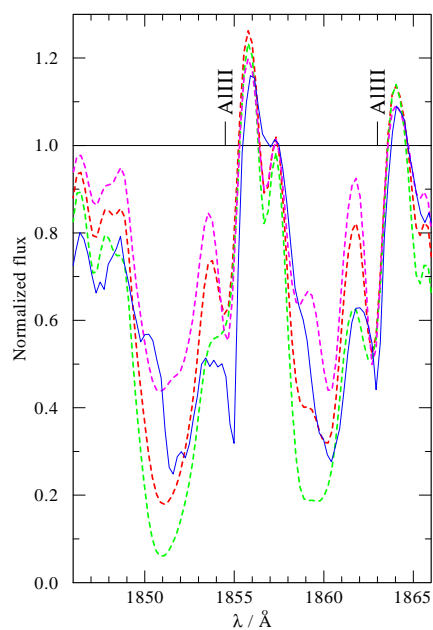, width=0.3\textwidth}
\caption{\footnotesize \AlIII ~resonance lines $\lambda 1855$ and $\lambda 1863\AA$, employed for the 
$\dot{M}$ estimation in Vela~X-1. We show the observation (solid blue line), the 
best-fit model ($\dot{M}=10^{-6.2} \, M_{\odot}/yr$, $D=11$, red dashed line), a 
model with higher $\dot{M}$ ($\dot{M}=10^{-5.8} \, M_{\odot}/yr$, $D=2$, green 
dashed line), and a model with lower $\dot{M}$ ($\dot{M}=10^{-6.3} \, 
M_{\odot}/yr$, $D=20$, pink dashed line).  }
\label{fig:Mdot_Vela}
\end{figure}

We derived $\dot{M}$ and $D$ from the \AlIII ~resonance lines $\lambda 1855$ and $\lambda 1863\AA$. As we
can see in Fig.~\ref{fig:Mdot_Vela}, the variation of $\dot{M}$ (and consequently $D$) directly affect these
lines. Higher (lower) $\dot{M}$ enhances (reduces) the density of the stellar wind, producing too strong (weak)
absorption. \\

\begin{figure}
\centering
\epsfig{file=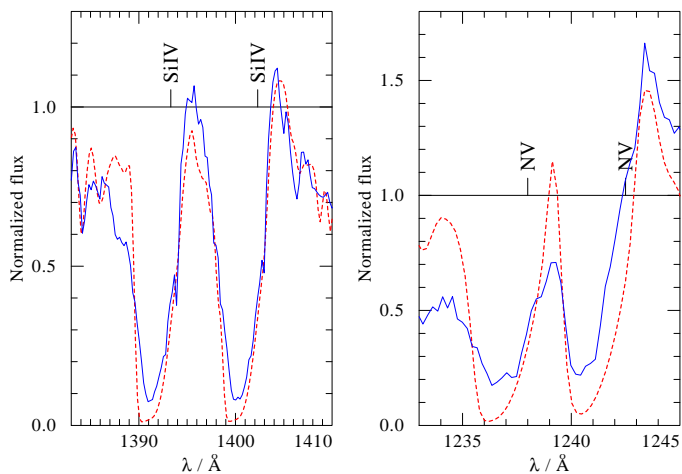, width=0.48\textwidth}
\caption{\footnotesize \SiIV ~and \NV ~resonance lines in Vela~X-1. While the observations show slight desaturation, 
all the models within a reasonable parameter space around the best-fit model produce saturated lines. }
\label{fig:desaturated}
\end{figure}

Unfortunately, other resonance lines available in the spectrum (\NV , \CIV ~and \SiIV ) are saturated in
the models within a reasonable range of parameters around the best-fit, and consequently are not
suitable for the $\dot{M}$ diagnosis. Interestingly, in contrast to the models, the \NV ~and \SiIV
~resonance lines are slightly desaturated in the observations (see Fig.~\ref{fig:desaturated}). The origin of this
phenomenon might be related to the presence of optically thick clumps (macroclumping), which directly
affects the mass-loss rate estimations \citep{2007A&A...476.1331O, 2012A&A...541A..37S}.
Undoubtedly, its study deserves further investigation, which is beyond the scope of this work. \\

\begin{figure}
\centering
\epsfig{file=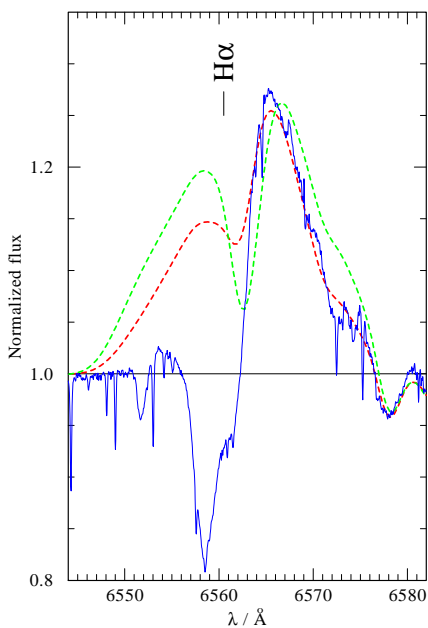, width=0.3\textwidth}
\caption{\footnotesize H$\alpha$ ~used the estimation of the parameter $\beta$ in Vela~X-1. We present the observation (solid blue line), the best-fit model ($\beta=1.0$, red dashed line), and a model with $\beta=0.5$ (green dashed line), as proposed by \cite{2015arXiv150701016M}.  }
\label{fig:beta_Vela}
\end{figure}

Based on the X-ray data analysis, \cite{2015arXiv150701016M} have suggested that 
the velocity law with the parameter $\beta=0.5$ fits better with the 
X-ray light curve of the system in near-to-eclipse phases. However, a satisfying fit is not 
possible when we assume $\beta=0.5$. We have tried models using $\beta=0.5$ and
adapting $\dot{M} \sqrt{D}$ in order to fit H$\alpha$. However, as shown in Fig.~\ref{fig:beta_Vela}, 
H$\alpha$ in our observation is not compatible with $\beta=0.5$. As we mentioned above, H$\alpha$ might 
suffer from important variability along the 
orbit. Moreover, the X-ray irradiation from the neutron star might produce 
variations in the stellar wind. In our opinion, this might be the cause of the apparent disagreement between 
the conclusions extracted from the X-rays and the optical wavebands.   \\

The chemical composition was estimated following the same approach as it was 
done for IGR~J17544-2619. Interestingly, we found again indications of chemical 
evolution in the star, given the moderate overabundance of He and N, together 
with the underabundance of C and O (see Table~\ref{tab:chemic}).  \\

\begin{figure}
\centering
\epsfig{file=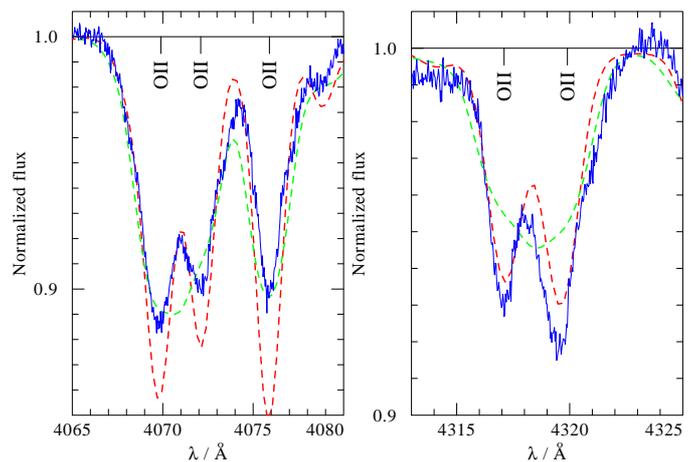, width=0.48\textwidth}
\caption{\footnotesize Example of unblended lines in the spectrum of Vela X-1. 
In red we plot a model with $\varv_\text{rot}\,\sin i = 56$~km/s. In green we plot a 
model with $\varv_\text{rot}\,\sin i = 116$~km/s.  }
\label{fig:rot_Vela}
\end{figure}

We adopted the value of $\varv_\text{rot}\,\sin i = 56$~km/s derived by 
\cite{2010MNRAS.404.1306F}. Previous estimations pointed to much higher values 
around 115~km/s \citep{1995A&A...299...79Z, 1997MNRAS.284..265H}, but such a 
high rotational velocity is not compatible with some of the lines that we see 
unblended in the optical observation (see Fig.~\ref{fig:rot_Vela}). The 
rotational velocity directly affects the estimation of the neutron star mass 
($M^\text{NS}_{\textsc{Vela\,X-1}}$) from radial velocity curves, as shown by 
\cite{2012A&A...539A..84K}. If $\varv_\text{rot}\,\sin i = 56$~km/s, it is feasible that 
$M^\text{NS}_{\textsc{Vela\,X-1}} \sim 1.5 \, M_{\odot}$, close to the canonical value ($1.4\,M_{\odot}$),
instead of a high mass neutron star 
$M^\text{NS}_{\textsc{Vela\,X-1}} \gtrsim 1.8 \, M_{\odot}$, as suggested by other 
authors \citep[e.g.][]{2003A&A...401..313Q, 2001A&A...377..925B}. \\

To summarise, our new analysis of Vela~X-1 is in 
broad agreement with previous studies of this system. 
We find a rather low stellar wind velocity, while $\dot{M}$ 
is typical for the stars of its spectral type. Like other 
studies, we note spectral line variability in dependence 
with orbital phase, and attribute it to the influence of 
the X-rays and the compact object on the stellar wind.  \\

The final physical parameters of the the two sources obtained 
in this work are shown in Table~\ref{tab:params_fit}.

\section{Discussion \label{sec:disc}}

\subsection{Wind-fed accretion \label{subsec:wind}}
In SFXTs and SGXBs, the X-ray emission is powered by the accretion of matter 
from the donor's wind onto the compact object. The efficiency of the conversion 
of the potential energy into X-ray luminosity depends on many factors including 
the properties of the stellar wind, the properties of the compact object and the 
orbital separation.  \\
\begin{table}
\begin{tabular}{P{0.14\textwidth} | P{0.14\textwidth} | P{0.14\textwidth} }
\toprule
Parameters & J17544-2619 & Vela~X-1 \\
\midrule
$P_\text{orb}$~(d)  & $4.9^\text{a}$ & $8.964357^b$  \\
$P_\text{spin}$~(s)   & $71.49^c$, $11.58^d$ & $283.532^b$  \\
$a \, \sin i$~(lt-s) & - & $113.89^b$  \\
$i$ (deg) & - & $>73^e$ \\
$a$~($10^{12}$cm) & $2.6^i$ & $3.5^j$ \\
$a$~($R_\star$) & $1.9$ & $1.8$ \\
$B$~($10^{12}$G)   & $1.45^f$ & $2.6^g$  \\
$\varv_\text{wind}$~(km/s) & $789^h$  &  $264^h$ \\
$\varv_\text{orb}$~(km/s) & $386^k$  &  $281^k$ \\
\bottomrule
\end{tabular}
\caption{\footnotesize \label{tab:params_disc} Parameters used in 
Sect.~\ref{sec:disc}. \\ References: (a)~\cite{2009MNRAS.399L.113C} 
(b)~\cite{2008A&A...492..511K} (c)~\cite{2012A&A...539A..21D} 
(d)~\cite{2015A&A...576L...4R} (e)~\cite{1995A&A...303..483V}  
(f)~\cite{2015MNRAS.447.2274B} (g)~\cite{2002A&A...395..129K} 
(h)~this work. (i)~From $P_\text{orb}$, total mass of the system and the 3rd Kepler's law.
(j)~From $a \, \sin i$, and the average $\langle \sin i \rangle = 0.985$ over $i>73^{\circ}$.
(k)~Assuming a circular orbit. }
\end{table}

The most efficient way of producing X-rays is the so called direct accretion: 
the stellar wind that is gravitationally captured by the neutron star free-falls 
onto the compact object. The expected luminosity is close to the accretion 
luminosity $L_\text{acc}$. The following equations contain the most relevant 
parameters in this regime:   
\begin{subequations}
\begin{align}
R_a &= \frac{2GM_\text{NS}}{\varv_\text{rel}^2} \label{eq:ra} \\
f_a &= \frac{R_a^2}{4a^2} \label{eq:fa} \\
L_\text{acc} &= f_a\,\frac{GM_\text{NS}\dot{M}}{R_\text{NS}} \label{eq:lacc}
\end{align}
\end{subequations}
where $R_a$ is the accretion radius (also called Bondi radius), that is to say, 
the maximum distance to the neutron star where the stellar wind is able to avoid 
falling onto the compact object; $G$ is the gravitational constant; $M_\text{NS}$ is 
the mass of the neutron star, which in this work is hereafter assumed to be 
the canonical value $1.4\,M_{\odot}$; $R_{NS}$ is the radius of the neutron star, which in this work 
is henceforward assumed to be $12$~km \citep{2014ApJ...784..123L};
$\varv_\text{rel}$ is the velocity of the wind relative to the 
neutron star; $f_a$ is the fraction of stellar wind that is gravitationally 
captured by the neutron star; $a$ is the orbital distance and $L_\text{acc}$ is the 
accretion luminosity, namely, the luminosity that would arise if the whole 
potential energy of the accreted matter is eventually transformed in X-ray 
luminosity. \\

For IGR~J17544-2619, using the results of our spectral fitting, the data shown in
Table~\ref{tab:params_disc} and assuming a circular orbit, we obtain from Eq.~\ref{eq:lacc}: 
$$L_\text{acc} = 1.4 \cdot 10^{36} \, \text{erg/s}$$
The value of $L_\text{acc}$ is 1-2 orders of magnitude higher than the luminosity 
that the source exhibits most of the time:  
$L_\text{X} < 5\cdot10^{34}$~erg/s \citep{2015AdSpR..55.1255B}.   
Most likely, some inhibition mechanism is acting in IGR~J17544-2619 
\citep{2014MNRAS.439.2175D, 2008ApJ...683.1031B}. \\

As a possible explanation for the variability of IGR~J17544-2619 and its lower-than-expected
luminosity at quiescence,
\cite{2008ApJ...683.1031B} discussed the application of their model to the light curve of an outburst 
observed by \textit{Chandra}. This theoretical framework describes the mechanisms for the
inhibition of the accretion according to the relative size of the spheres 
defined by $R_\text{a}$, $R_\text{M}$ and $R_\text{co}$; where $R_\text{a}$ is the already
defined accretion radius, $R_\text{M}$ is the magnetospheric radius (location where the
pressure exerted by the gas equals the local magnetic pressure), and $R_\text{co}$
is the co-rotation radius (location where the angular velocity of the neutron star 
equals the Keplerian velocity). These radii, in turn, depend on: $\dot{M}$, $\varv_\text{rel}$, magnetic 
moment of the neutron star ($\mu$), orbital separation ($a$) and $P_\text{spin}$. \\

For simplicity,
the orbital velocity of the neutron star and the eccentricity are not considered 
in the model by \cite{2008ApJ...683.1031B}. That is to say, it is assumed that 
$e=0$ and $\varv_\text{wind} \simeq \varv_\text{rel}$, where $\varv_\text{wind}$ is the stellar wind
velocity in the position of the neutron star.
We note that when the stellar wind velocity is not very high, this assumption might not be 
accurate. Indeed, the orbital velocity ($\varv_\text{orb}$) in Vela~X-1
is very similar to $\varv_\text{wind}$ (see Table~\ref{tab:params_disc}). 
In IGR~J17544-2619, the orbital velocity is around the half of
the stellar wind velocity. 
Despite these simplifications, the model provides significant
insight on the explanation of the qualitative behavior of the sources 
with regard to their persistence or 
variability, as it is shown next in this section. A more accurate approach considering
eccentric orbits and the orbital velocity of the compact object would be an important
advance in the model, but it is out of the scope of this paper.   \\
 
Nowadays, the tentative estimations of the spin period in IGR~J17544-2619 ($P_\text{spin}=71.49$s 
by \cite{2012A&A...539A..21D} and alternatively $P_\text{spin}=11.58$s by 
\citealt{2015A&A...576L...4R}), along with the stellar wind parameters derived in 
this work, permit to discuss the application of the model by \cite{2008ApJ...683.1031B} from a new 
perspective. The rest of parameters required for this section are shown in 
Table~\ref{tab:params_disc}. Using those values, we can elaborate
diagrams $\varv_\text{wind}$-$\dot{M}$ and $\varv_\text{wind}$-$P_\text{spin}$, where
the different accretion regimes occupy different domains of the space of parameters.
These domains directly arise from the Eq.~25-28 by \cite{2008ApJ...683.1031B}. \\

\begin{figure}
\centering
\epsfig{file=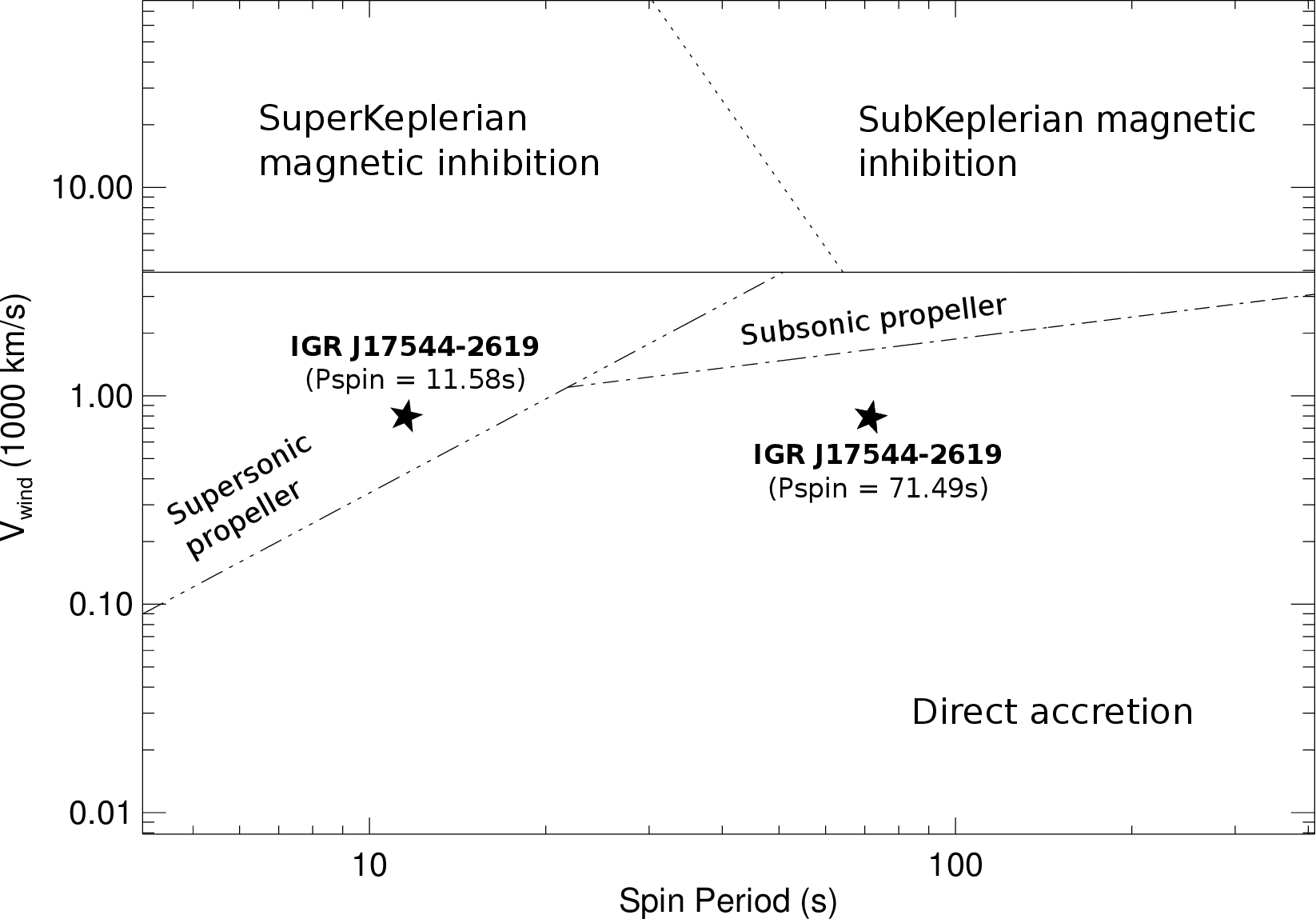, width=0.48\textwidth}
\caption{\footnotesize Position of IGR~J17544-2619 in the 
$\varv_\text{wind}$-$P_\text{spin}$ diagram for the two tentative estimations of $P_\text{spin}$.
Equations 25, 26, 27 and 28 by \cite{2008ApJ...683.1031B} are represented by a solid, dotted, triple-dot-dashed and dot-dashed lines respectively.}
\label{fig:alterBozzo_17544}
\end{figure}

In Fig.~\ref{fig:alterBozzo_17544} we show the position of IGR~J17544-2619 in the 
diagram $\varv_\text{wind}$-$P_\text{spin}$ for the two currently available tentative 
estimations of the $P_\text{spin}$. The source lie in the direct accretion 
regime for $P_\text{spin}=71.49$s, and in the supersonic propeller regime for 
$P_\text{spin}=11.58$s. Hence, the shortest $P_\text{spin}=11.58$s matches 
better with the X-ray behavior of the source and its likelihood of staying in an 
inhibited accretion regime. \\

\begin{figure}
\centering
\epsfig{file=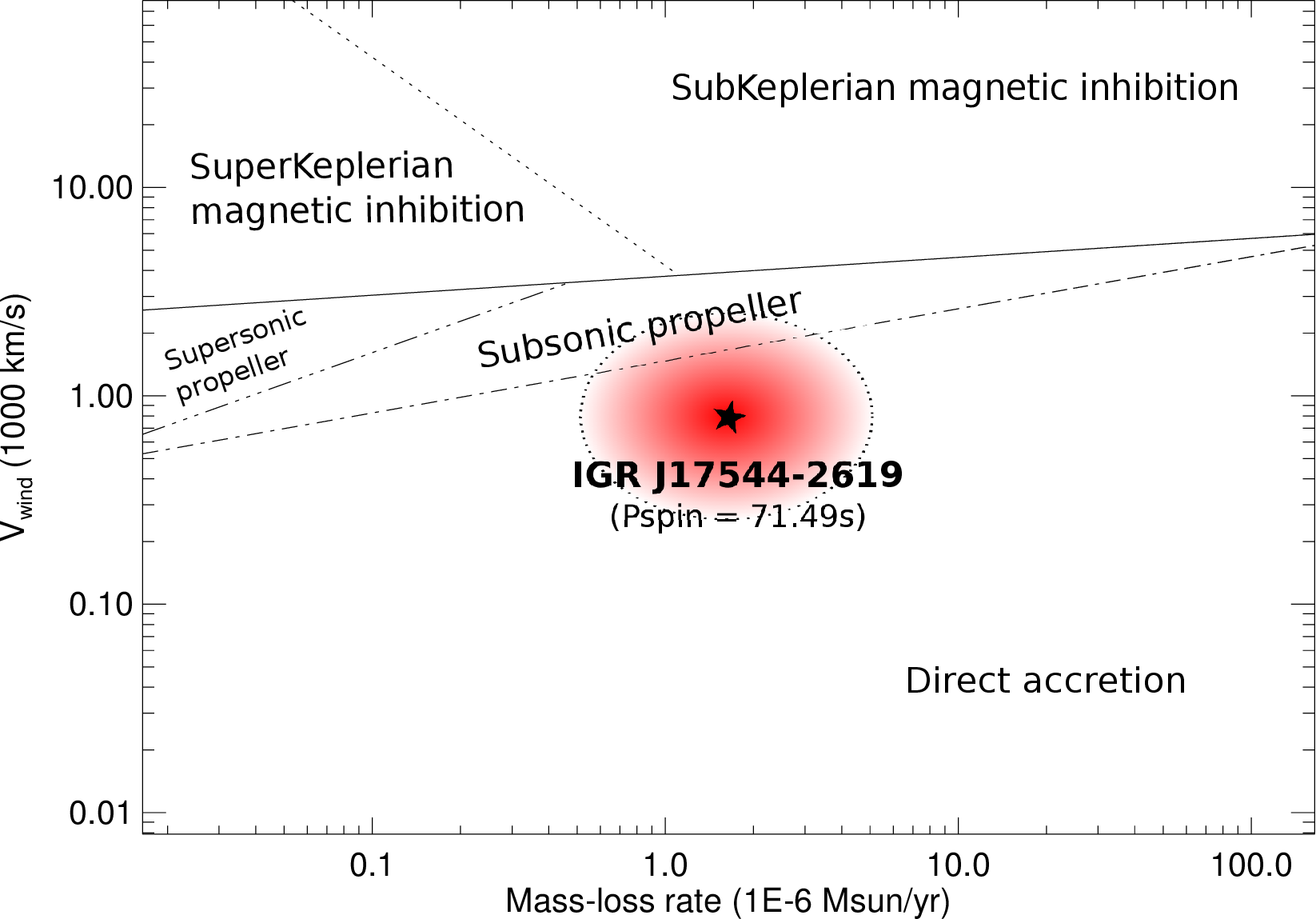, width=0.48\textwidth}
\epsfig{file=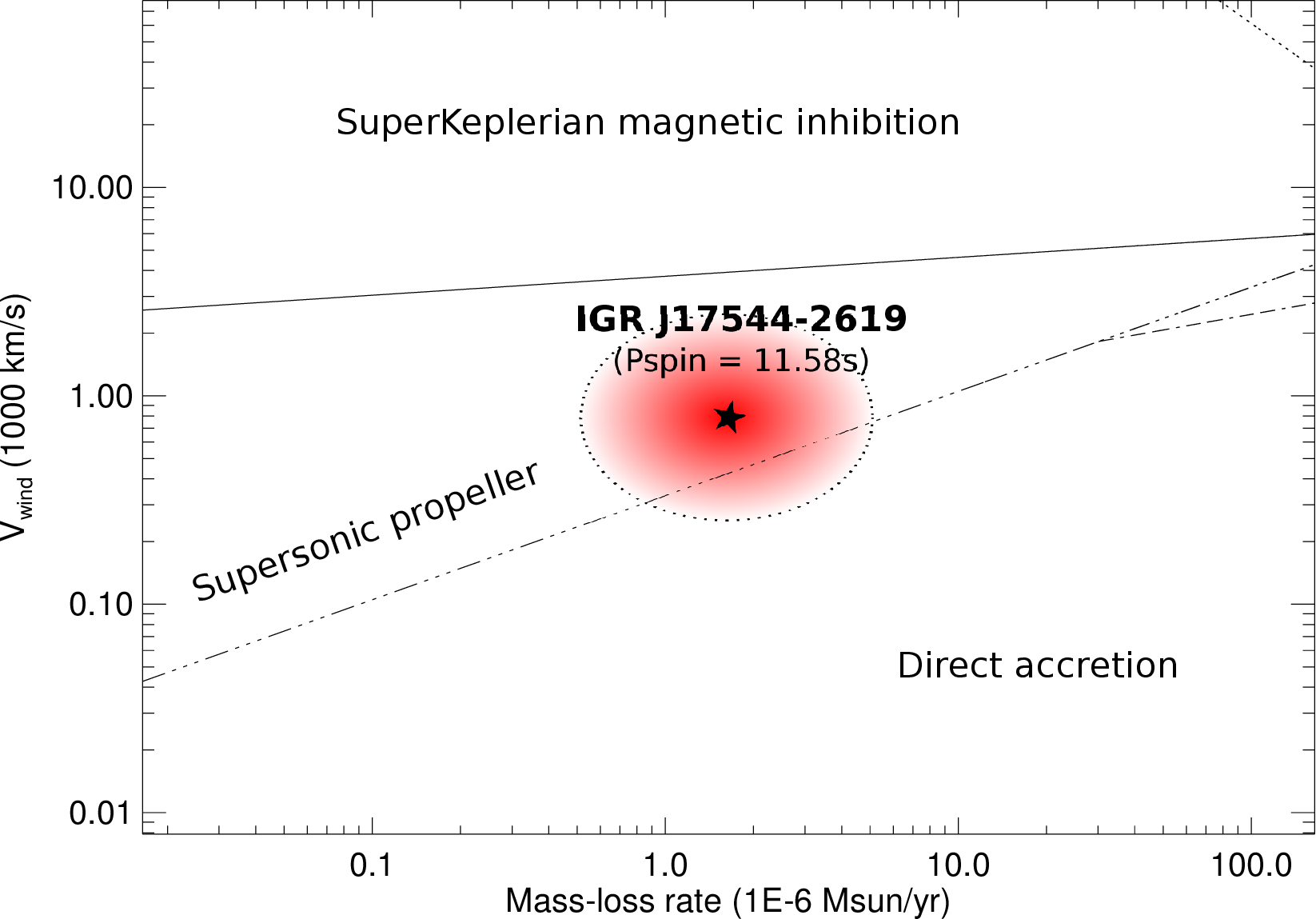, width=0.48\textwidth}
\caption{\footnotesize Position of IGR~J17544-2619 in the 
$\varv_\text{wind}$-$\dot{M}$ diagram. Upper panel: diagram calculated using 
$P_\text{spin}=11.58$s. Lower panel: diagram calculated using 
$P_\text{spin}=71.49$s. The dashed line encircles the space within one order of magnitude 
of $\varv_\text{wind}$ and $\dot{M}$. 
Equations 25, 26, 27 and 28 by \cite{2008ApJ...683.1031B} are represented by a solid, dotted, triple-dot-dashed and dot-dashed lines respectively. }
\label{fig:Bozzo_17544}
\end{figure}

In Fig.~\ref{fig:Bozzo_17544} we show the location of IGR~J17544-2619 in the 
diagram $\varv_\text{wind}$-$\dot{M}$. It is important to note that the position
of the system in this diagram is not a fixed point due to the the intrinsic variability of 
the velocity and local density of the stellar wind in hot massive stars. Thus,
we have plotted a red region in Fig.~\ref{fig:Bozzo_17544} showing a variability
of one order of magnitude in $\varv_\text{wind}$ and $\dot{M}$. That is to say,
the maximum $\dot{M}$ and $\varv_\text{wind}$ in the encircled region is ten times higher than the 
minimum $\dot{M}$ and $\varv_\text{wind}$. 
Such a variability is fully plausible, as demonstrated by hydrodynamical simulations of 
radiatively driven stellar winds \citep[e.g.][]{1997A&A...322..878F}.
These clumps of higher density, 
intrinsic to stellar winds of hot stars, 
are sometimes invoked to explain the X-ray 
variability of HMXBs \citep{2012MNRAS.421.2820O}.
As we can see in Fig.~\ref{fig:Bozzo_17544}, the encircled region intersects regimes of direct
accretion and inhibited accretion.
Hence, it is possible that in objects like IGR~J17544-2619, the abrupt changes in the
wind density may lead to the switching from one accretion regime to the other. 
Moreover, besides the clumping of the stellar wind, the eccentricity of the orbit 
($e<0.25$) would lead to additional variations in the 
orbital separation (and consequently in $\varv_\text{wind}$ and the density of the medium), 
which reinforce the intrinsic variability of the stellar wind and its capability
to lead to transitions across regimes.  \\

Considering an alternative explanation for the X-ray variability of 
IGR~J17544-2619, \cite{2014MNRAS.439.2175D} invoked the quasi-spherical 
accretion model by \cite{2012MNRAS.420..216S}. However, if the spin period is 
actually as short as $71.49$s or $11.58$s, the condition of a slowly rotating 
pulsar, i.e. R$_\text{M} \ll \text{R}_\text{co}$ (where R$_\text{M}$ is the magnetospheric radius
and R$_\text{co}$ the co-rotation radius), assumed by this approach, would be debatable. 
Even though it raises doubts about the feasibility of applying this model, it 
cannot be ruled out until the spin period and the magnetic field of the neutron 
star are firmly constrained.   \\

\begin{figure}
\centering
\epsfig{file=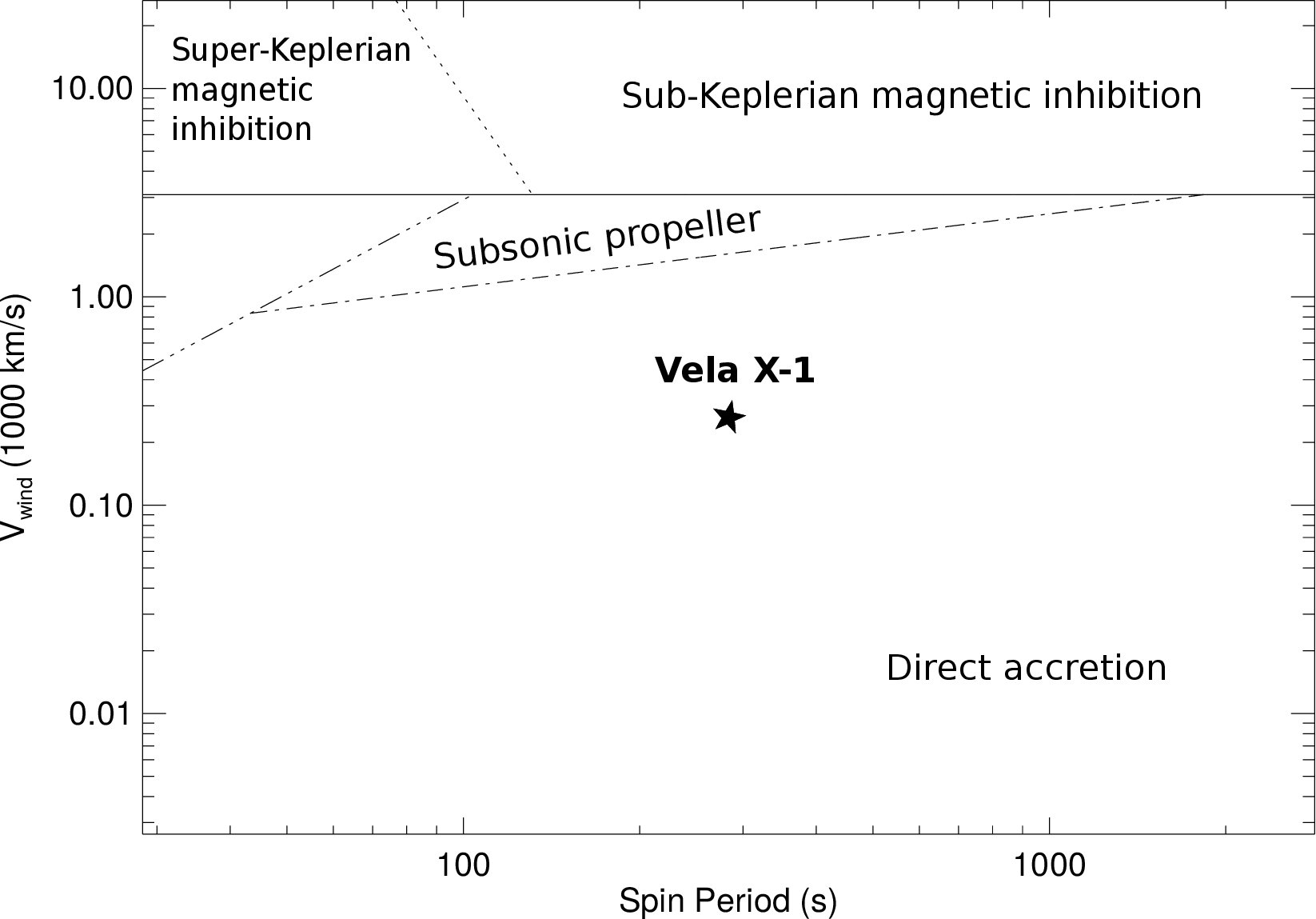, width=0.48\textwidth}
\caption{\footnotesize Position of Vela~X-1 in the 
$\varv_\text{wind}$-$P_\text{spin}$ diagram. 
The dashed line encircles the space within one order of magnitude 
of $\varv_\text{wind}$ and $\dot{M}$. 
Equations 25, 26, 27 and 28 by \cite{2008ApJ...683.1031B} are represented by a solid, dotted, triple-dot-dashed and dot-dashed lines respectively.}
\label{fig:alterBozzo_Vela}
\end{figure}

\begin{figure}
\centering
\epsfig{file=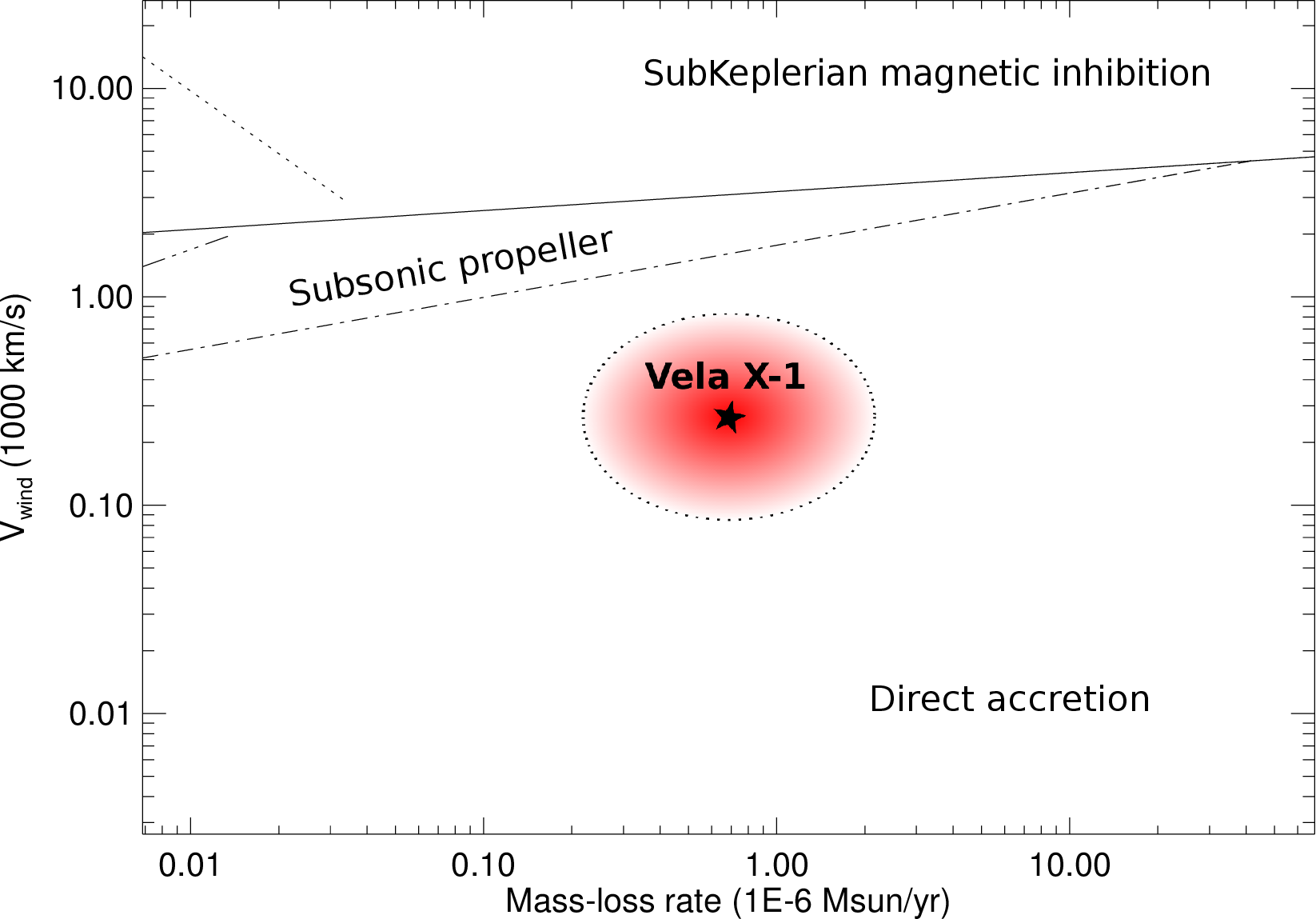, width=0.48\textwidth}
\caption{\footnotesize Position of Vela~X-1 in the $\varv_\text{wind}$-$\dot{M}$ diagram.  
Equations 25, 26, 27 and 28 by \cite{2008ApJ...683.1031B} are represented by a solid, dotted, triple-dot-dashed and dot-dashed lines respectively.}
\label{fig:Bozzo_Vela}
\end{figure}

In the case of Vela~X-1, we can see in Fig.~\ref{fig:alterBozzo_Vela} and ~\ref{fig:Bozzo_Vela}, 
the source is well in the middle of the zone where direct accretion is expected.
Hence, more extreme density or velocity jumps 
would be required to trigger any change of accretion regime. These extreme 
jumps are also plausible, but much more unlikely. However, they might sporadically 
occur and lead to a sudden decrease of the luminosity in Vela~X-1. \\ 

Using the parameters shown in Table~\ref{tab:params_disc} and Eq.~\ref{eq:lacc}, 
we obtain $L_\text{acc} = 8.7 \times 10^{36}$~erg/s for Vela~X-1. 
The average X-rays luminosity of the source is
$\langle L_\text{X} \rangle \simeq 4.5 \times 10^{36}$ \citep{1999ApJ...525..921S}.
More specifically, $L_\text{acc} \simeq 0.5 \times \langle L_\text{X} \rangle$. This means
that there is a good agreement between $L_\text{acc}$ and $L_\text{X}$,
which implies that the direct accretion scenario can describe the way 
that matter is accreted in Vela~X-1. \\

The framework of different accretion regimes described by 
\cite{2008ApJ...683.1031B} is able to explain why IGR~J17544-2619 is prone to 
show a high X-ray variability and inhibited accretion (assuming the shortest 
$P_\text{spin}=11.58$s), and Vela~X-1 is persistently very luminous in the 
X-rays. As exposed in Fig.~\ref{fig:Bozzo_17544} and \ref{fig:Bozzo_Vela}, 
the required variability in the stellar wind for a transition in the accretion 
regime is far lower in IGR~J17544-2619 than in Vela~X-1. The main ingredients 
that make the sources so different are the $P_\text{spin}$ (shorter in 
IGR~J17544-2619), and the $\varv_\text{wind}$ (larger in IGR~J17544-2619). \\

We may conjecture whether this theoretical framework can be applied to other SGXBs and SFXTs.
Unfortunately there are not many sources where we can find complementary studies including 
dedicated analysis of the stellar wind, orbital parameters and neutron star parameters. 
The studies of the stellar wind are specially scarce. Besides the two sources analysed
in this work, there are at least four where a comparable amount of information is 
available in the literature. They are IGR~J11215-5952, GX~301-2, X1908+075 and OAO 1657-415.
We show the diagrams $\varv_\text{wind}$-$P_\text{spin}$ and $\varv_\text{wind}$-$\dot{M}$
for these sources in Appendix~\ref{app:others}. Again, the diagrams seem to qualitatively explain the
behavior of the systems. GX~301-2, X1908+075 and OAO~1657-415
are persistent SGXBs, and they occupy regions of highly likelihood of persistent emission in the
diagrams. In contrast, the likelihood of regime transitions in IGR~J11215-5952 is much higher. \\

IGR~J11215-5952 is a system with very large eccentricity and long orbital period \citep{2009ApJ...696.2068R}.
It shows recurrent flares with a period of $\sim 330$d \citep{2006A&A...450L...9S}. Its high
variability leads to its classification as a SFXT, even though the predictability of the flares 
is not a common feature in the rest of SFXTs. 
\cite{2007A&A...476.1307S} proposed that the recurrent flares might be
explained by an additional equatorial component of the stellar wind
combined with the highly eccentric orbit. In Figure~\ref{fig:Bozzo_11215} we can see that a
moderate clumpiness would lead to frequent transition regimes, and hence
it would be expected a very high X-rays variability. However, the diagram shown
in Figure~\ref{fig:Bozzo_11215} is calculated assuming a circular orbit, which
is not accurate for IGR~J11215-5952. In this source,
the high eccentricity of the system might be a more important factor than 
the clumpiness of the wind,
and the transition into the direct accretion regime might be likely only during near-periastron passages, 
producing periodic outbursts. \\

Regarding other systems, the framework used here might encounter problems to explain the 
behavior of other SFXTs with larger $P_\text{spin}$ such as IGR~J16418-4532 ($P_\text{spin}=1212$s,
\citealt{2012MNRAS.420..554S}) and IGR~J16465-4507 ($P_\text{spin}=228$s, \citealt{2005A&A...444..821L}).
The estimation of the stellar wind parameters in these systems will be 
very useful to measure the extent of the applicability of the model by \cite{2008ApJ...683.1031B} 
explaining the dichotomy between SGXBs and SFXTs. Moreover, studies of the X-rays absorption
might provide an additional perspective on the issue. \cite{2015A&A...576A.108G}
studied a sample of SGXBs and SFXTs using XMM-Newton and it was observed that the SGXBs 
included in the sample were in general more absorbed than the SFXTs.
This may suggest a more intense interaction of the X-rays radiation with the stellar wind, or, 
alternatively, that the neutron star orbits a more dense medium in SGXBs due to 
a closer orbit or a slower stellar wind of the donor. \\ 

Finally, we can compare the $\varv_\infty$ and the $\varv_\text{esc}$ that we obtain 
from the fits. In this regard, \cite{1995ApJ...455..269L} collected a
large dataset from hot stars with radiatively driven winds, and concluded that the ratio
$\varv_\infty/\varv_\text{esc}$ steeply decreases from $\sim 2.6$ to $\sim 1.3$ when going
from high to low $T_\text{eff}$ at a point near $T_\text{eff} \simeq 21000$~K, corresponding
to spectral type around B1. According to \cite{1999A&A...350..181V}, this drop is caused 
by a decrease in the line acceleration of \FeIII ~in the subsonic part of the wind. In our case 
we have (see Table~\ref{tab:params_fit}): 
\begin{itemize}
\item IGR~J17544-2619 (O9.5I): $\varv_\infty / \varv_\text{esc} = 2.4_{-0.5}^{+0.7}$
\item Vela~X-1 (B0.5I): $\varv_\infty / \varv_\text{esc} = 1.6_{-0.4}^{+0.8}$
\end{itemize}
These values follow the trend observed and described by \cite{1995ApJ...455..269L}. 
We suggest that it might be the reason why IGR~J17544-2619 shows higher $\varv_\infty$ than Vela~X-1. 
The action of the X-rays can also make an important impact in the velocity of the
stellar wind, as shown by \cite{2014PASJ...66...34K}. However, this effect is probably local, 
since we do not observe important differences in the terminal velocity between eclipsing and 
non-eclipsing orbital phases in Vela~X-1. Secondary features like asymmetries or additional
absorption components in the spectral lines, which might be related to the effect of the
X-rays in the stellar wind, are described and discussed below in Sect.~\ref{sec:asym}.   
 
\subsection{Evolutionary tracks}
\begin{figure}
\centering
\epsfig{file=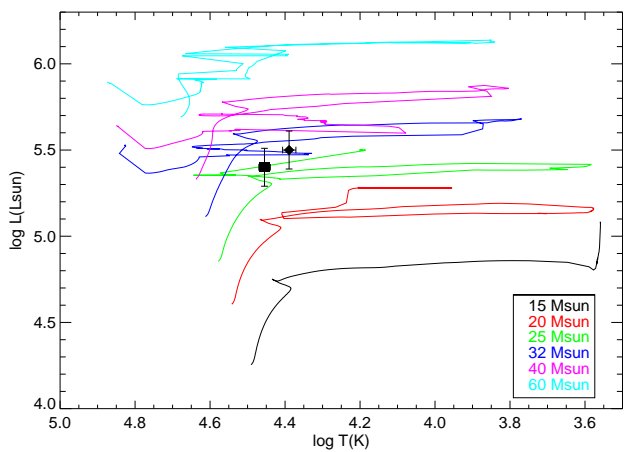, width=0.48\textwidth}
\caption{\footnotesize Evolutionary tracks from the Geneva Stellar Models with 
solar abundances and rotation. The positions of IGR~J17544-2619 (square) and 
Vela~X-1 (diamond) are overplotted. }
\label{fig:HRD}
\end{figure}

In Fig.~\ref{fig:HRD} we show the position of Vela~X-1 and IGR~J17544-2619 in 
the Hertzprung-Russell Diagram (HRD), and the evolutionary tracks from the Geneva 
Stellar Models \citep{2012A&A...537A.146E}. The two stars lie on the theoretical 
track of a star with initial mass $\sim 25-30\,M_{\odot}$. In IGR~J17544-2619 the spectroscopic 
mass obtained from the fits is compatible with the evolutionary mass. Vela~X-1 shows certain overluminosity, since
its spectroscopic mass is lower than the evolutionary mass. Nevertheless, 
the mass of the star obviously decreases
along its lifetime due to the stellar wind and possible mass transfer episodes. These phenomena might 
have been stronger or longer in Vela~X-1 compared to IGR~J17544-2619.   \\

The overabundance of helium and nitrogen arising from the fits in the two stars
might trigger an increase in luminosity following the scaling relation $L 
\propto \mu^\alpha$, where $\mu$ is the average mean molecular weight and $\alpha 
>1$ \citep{1992A&A...265L..17L}. Then, we expect certain overluminosity in both 
sources. However, as already mentioned, the overluminosity is more noticeable in Vela~X-1. In all,
the sources seem to be in a different evolutionary stage or to have experienced 
a different evolutionary history. \\

The chemical evolution of the donors might have been driven by episodes of 
important mass transfer in the past, given the close orbits of the systems, enhancing the helium and nitrogen abundances due to the accretion of chemically enriched material \citep{2012ARA&A..50..107L}. Moreover, Roche-lobe overflow stages induce important spin-up in the mass gainer \citep{1981A&A...102...17P}, inducing further chemical enrichment because of rotational mixing. This scenario is supported by the observation of other HMXBs where indications of nitrogen enhancement are also observed \citep{2014A&A...566A.131G}.  \\

\subsection{Asymmetries in spectral lines of Vela~X-1 \label{sec:asym}}
\begin{figure}
\centering
\epsfig{file=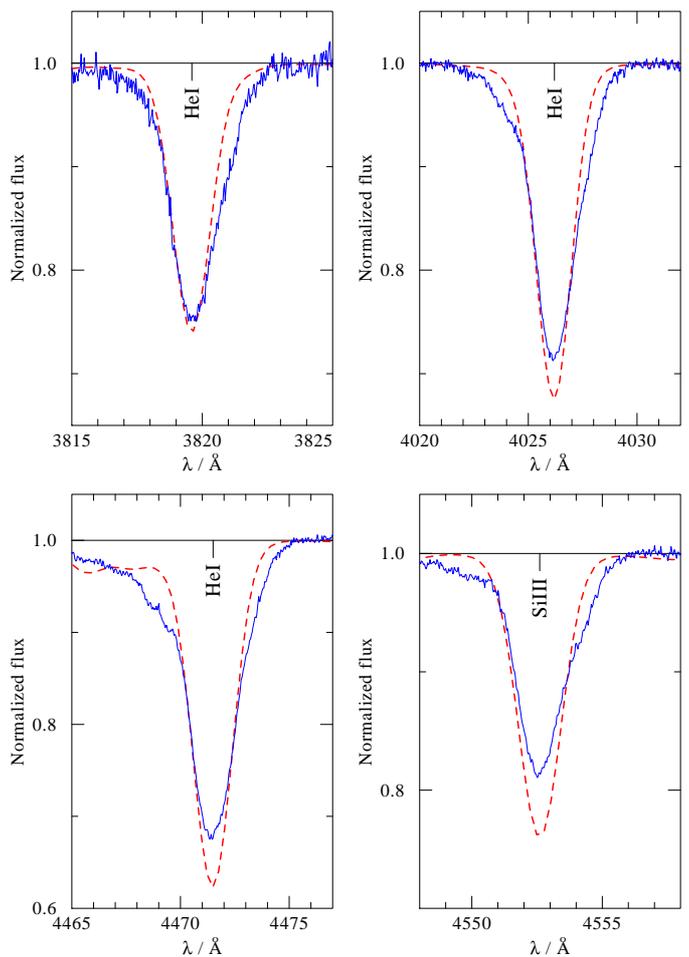, width=0.48\textwidth}
\caption{\footnotesize Example of four spectral lines showing notable 
asymmetries: \HeI ~$\lambda 3820, 4026, 4471 \, \AA$, and \SiIII ~$\lambda 4553 \, \AA$.}
\label{fig:asym}
\end{figure}

Some of the lines in the spectrum of Vela~X-1 show clear asymmetries that are 
not possible to reproduce with spherically symmetric models like PoWR (see Fig.~\ref{fig:asym}). This 
striking feature is specially noticeable for \HeI ~lines, but it is also observed 
in C, N, O or Si, whenever the lines are strong enough.  \\

Asymmetries in spectral lines were also reported by \cite{2015A&A...578A.107M} in hydrogen lines of the 
infrared spectrum of X1908+75, a SGXB. A natural explanation for the discrepancy between models and observations is the departure of the donor and/or the surrounding medium from the spherical 
symmetry. This departure may be triggered by tidally induced effects and the 
persistent X-ray irradiation of the stellar wind and the stellar surface. In 
this regard, \cite{2012A&A...539A..84K} showed that tidal effects would produce 
asymmetries in the line profiles.   \\

\begin{figure}
\centering
\epsfig{file=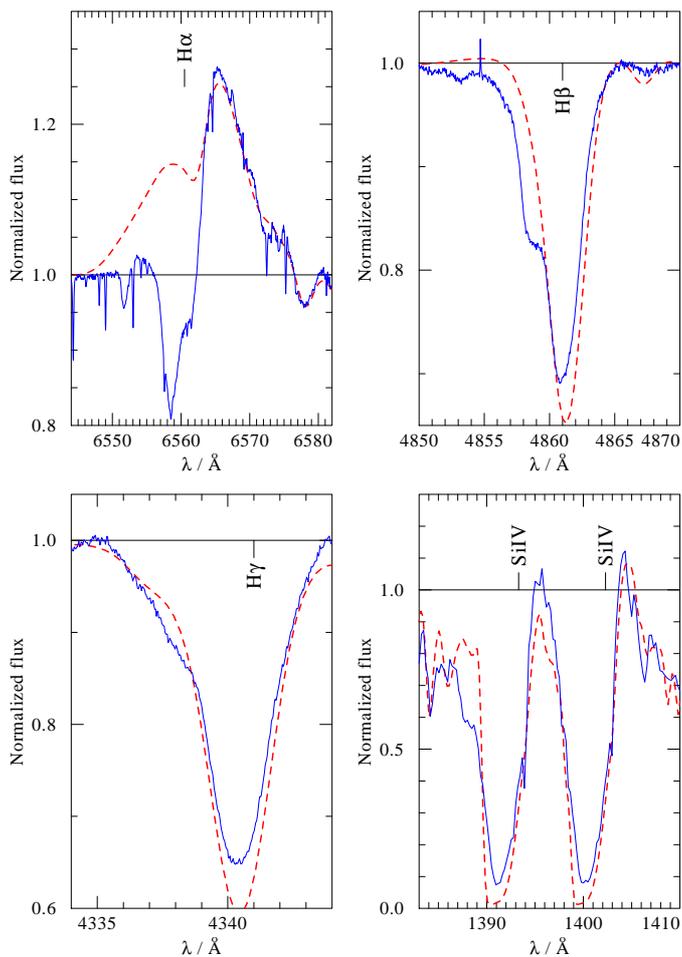, width=0.48\textwidth}
\caption{\footnotesize H$\alpha$, H$\beta$, H$\gamma$ and \SiIV ~$\lambda 1394,1403 \, \AA$. The observations (blue solid line) shows an additional blueshifted component that we are not able to reproduce with the models (red dashed line). }
\label{fig:blue_abs}
\end{figure}

The observed asymmetries might be related to the additional absorption that we observe in the blue part of
other important lines, with special attention to H$\alpha$, H$\beta$, H$\gamma$ and \SiIV ~$\lambda \, 1394,1403 \
\AA$ (see Fig.~\ref{fig:blue_abs}). Assuming that the absorption is produced by an independent component of matter
moving at certain velocity, it is striking that the involved velocities required for explaining such a blueshif are
different depending on the lines: $\sim 200-300$~km/s in H$\alpha$, H$\beta$ and H$\gamma$, $\sim 1000$~km/s in the
\SiIV ~resonance lines. \\ 

In any case, we note that these asymmetries and additional absorption features have not been
observed in IGR~J17544-2619. Hence, the physical cause at work is playing a 
significantly more important role in Vela~X-1 than in IGR~J17544-2619. This fact suggests that
the interaction of the X-ray source with the stellar wind might be fundamental for understanding these asymmetries, given that the X-rays are on average more intense in Vela~X-1. Indeed, if we compare the wind mechanic luminosity $L_\text{mech}=\dot{M}\,\varv_{\infty}^2 / 2$ to the X-ray luminosity $L_\text{X}$ we obtain:
\begin{itemize}
\item IGR~J17544-2619: $L_\text{mech} \simeq 10^{36}$~erg/s. That is to say, at least two orders of magnitude higher than the usual X-ray luminosity of the source.
\item Vela~X-1: $L_\text{mech} \simeq 10^{35}$~erg/s. Namely, about one order of magnitude lower than the X-ray luminosity of the source in quiescence.
\end{itemize}

Hence, there is a fundamental difference in the ratio $L_\text{mech}/L_\text{X}$. The 
X-rays are much more powerful with respect to the stellar wind in Vela~X-1 rather than in IGR~J17544-2619. 
We suggest that this fact might be related to the asymmetries that we observe 
in the spectral lines of Vela~X-1, but not in IGR~J17544-2619.

\section{Summary and conclusions \label{sec:conclusions}}
We have performed a detailed analysis of the donors of the HMXBs IGR~J17544-2619 
and Vela~X-1, using the code PoWR that computes models of hot stellar 
atmospheres. We found the luminosity, extinction, stellar mass, stellar radius, 
effective temperature, effective surface gravity, terminal velocity of the 
stellar wind, mass-loss rate, clumping factor, micro and macro-turbulent 
velocity, rotational velocity and chemical abundances. \\

The estimation of the above mentioned parameters has implications on other 
physical parameters of the system: the derived stellar radius of IGR~J17544-2619 
implies an upper limit in the eccentricity of the source: $e<0.25$. The 
rotational velocity derived for Vela~X-1 implies that the mass of the neutron star might be
$M^\text{NS}_{\textsc{Vela\,X-1}} \sim 1.5 \, M_{\odot}$, close to the canonical value ($1.4\,M_{\odot}$). \\

The donors of IGR~J17544-2619 and Vela~X-1 are similar in many of the parameters 
that physically characterise them and their spectrum. Moreover, they are also 
comparable in the eccentricity and orbital separation. 
However, in the context of accretion regimes described 
by \cite{2008ApJ...683.1031B}, their moderate differences in the stellar wind 
velocity and the $P_\text{spin}$ of the neutron star lead to a very different 
accretion regimes of the sources, which qualitatively explain their completely 
different X-ray behavior. After analysing other sources with sufficient
information available in the literature, we have observed that the same theoretical framework
is valid to qualitatively explain their X-ray behavior.
Further explorations addressing the estimation of the 
stellar wind properties of the donors in SGXBs and SFXTs, complemented with 
$P_\text{spin}$ measurements in SFXTs, will be necessary to confirm whether the 
conclusions exposed here can be extrapolated to additional members of these 
groups of HMXBs.    \\

In summary, this study shows that the wind terminal velocity 
play a decisive role in determining 
the class of HMXB hosting a supergiant donor. While low stellar wind velocity facilitates direct steady 
accretion in SGXBs, the high wind velocity and velocity jumps can easily shift the accretion mechanism 
from direct accretion to propeller regimes in SFXTs. This effects might be 
enhanced by other factors such as the eccentricity of the sources. We conclude that this is 
one of the mechanisms
responsible for these two major sub-classes of HMXBs with supergiant donors. \\

{\footnotesize \textit{Acknowledgments.} The work of AG-G has been supported by 
the Spanish MICINN under FPI Fellowship BES-2011-050874 associated to the 
project AYA2010-15431. 
T.S. is grateful for financial support from the Leibniz Graduate School for
Quantitative Spectroscopy in Astrophysics, a joint project of the Leibniz
Institute for Astrophysics Potsdam (AIP) and the Institute
of Physics and Astronomy of the University of Potsdam. 
This work has been partially supported by the 
Spanish Ministry of Economy and Competitiveness project numbers 
ESP2013-48637-C2-2P and ESP2014-53672-C3-3-P, 
the Generalitat Valenciana project number GV2014/088 
and the Vicerectorat d'Investigació, Desenvolupament i Innovació de la 
Universitat d'Alacant under grant GRE12-35. 
We wish to thank Thomas E. Harrison for his important 
contribution to the paper reducing the SpeX data. We also thank S. Popov for a very useful discussion.
The authors gratefully acknowledge the constructive comments on the paper given
by the anonymous referee.
SMN thanks the support of the Spanish Unemployment Agency, allowing her to continue her 
scientific collaborations during the critical situation of the Spanish Research System.
The authors acknowledge the help of the International Space Science Institute at Bern, 
Switzerland, and the Faculty of the European Space Astronomy Centre.
A.S. is supported by the Deutsche Forschungsgemeinschaft
(DFG) under grant HA 1455/26. 
Some of the data presented in 
this paper were obtained from the Multimission Archive at the Space Telescope 
Science Institute (MAST). STScI is operated by the Association of Universities 
for Research in Astronomy, Inc., under NASA contract NAS5-26555. Support for 
MAST for non-HST data is provided by the NASA Office of Space Science via grant 
NAG5-7584 and by other grants and contracts. This publication makes use of data 
products from the Two Micron All Sky Survey, which is a joint project of the 
University of Massachusetts and the Infrared Processing and Analysis 
Center/California Institute of Technology, funded by the National Aeronautics 
and Space Administration and the National Science Foundation. }

\bibliographystyle{aa}
\bibliography{bib}

\onecolumn
\appendix
\section{Other sources \label{app:others}}
In this appendix we show the diagrams that are further discussed in Sect.~\ref{sec:disc}, calculated from 
the available data in the literature, which is collected in Table~\ref{tab:app_params}.

\begin{table}[H]
\begin{tabular}{P{0.18\textwidth} | P{0.18\textwidth} | P{0.18\textwidth}  | P{0.18\textwidth} | P{0.18\textwidth}  | P{0.18\textwidth}}
\toprule
Parameters & J11215-5952 & GX~301-2 & X1908+075 & OAO 1657-415 \\
 & (SFXT) & (SGXB) & (SGXB) & (SGXB) \\
\midrule
$P_\text{orb}$~(d)  & $164.6^a$ & $41.508^d$ & $4.4007^g$ & $10.44812^i$ \\
$P_\text{spin}$~(s)  & $186.78^b$ & $685^e$ & $604.684^g$ & $38.2^j$ \\
$M_{\star}/M_{\odot}$  & $30^c$ & $43^e$ & $15^h$ & $14.3^k$ \\
$R_{\star} / R_{\odot}$ & $40^c$ & $70^e$ & $16^h$ & $24.8^k$ \\
$a$~($10^{12}$cm) & $27.2^l$ & $11.0^l$ & $2.4^l$ & $4.2^l$ \\
$a$~($R_\star$) & $9.8$ & $2.3$ & $2.1$ & $2.5^l$ \\
$B$~($10^{12}$G) & $1.45^n$ & $3.8^f$ & $1.45^n$ & $1.45^n$ \\
$\varv_\text{wind}$~(km/s) & $1128^c$ & $110^e$ & $235^h$ & $156^k$ \\
$\beta$ & $0.8^n$ & $1.75^e$ & $1.2^h$ & $0.9^k$ \\
\bottomrule
\end{tabular}
\caption{\footnotesize \label{tab:app_params} Parameters used in 
Appendix~\ref{app:others}. \\ References: (a)~\cite{2009ApJ...696.2068R} 
(b)~\cite{2007ATel..999....1S} (c)~\cite{2014A&A...562A..18L} 
(d)~\cite{1986ApJ...304..241S}
(e)~\cite{Kaper2006} (For $P_\text{spin}$, we used an intermediate value in the 
observed range $675\text{s} < P_\text{spin} < 700\text{s}$ in the 1974-2001 period). 
(f)~\cite{2004A&A...427..975K}
(g)~\cite{2004ApJ...617.1284L} (h)~\cite{2015A&A...578A.107M}
(i)~\cite{2008A&A...486..293B} (j)~\cite{1979ApJ...233L.121W}
(k)~\cite{2012MNRAS.422..199M}
(l)~From $P_\text{orb}$, total mass of the system and the 3rd Kepler's law. 
(n)~Not based in any estimation. Assumed as the same value as in IGR~J17544-2619. }
\end{table}

\begin{figure}[H]
\centering
\epsfig{file=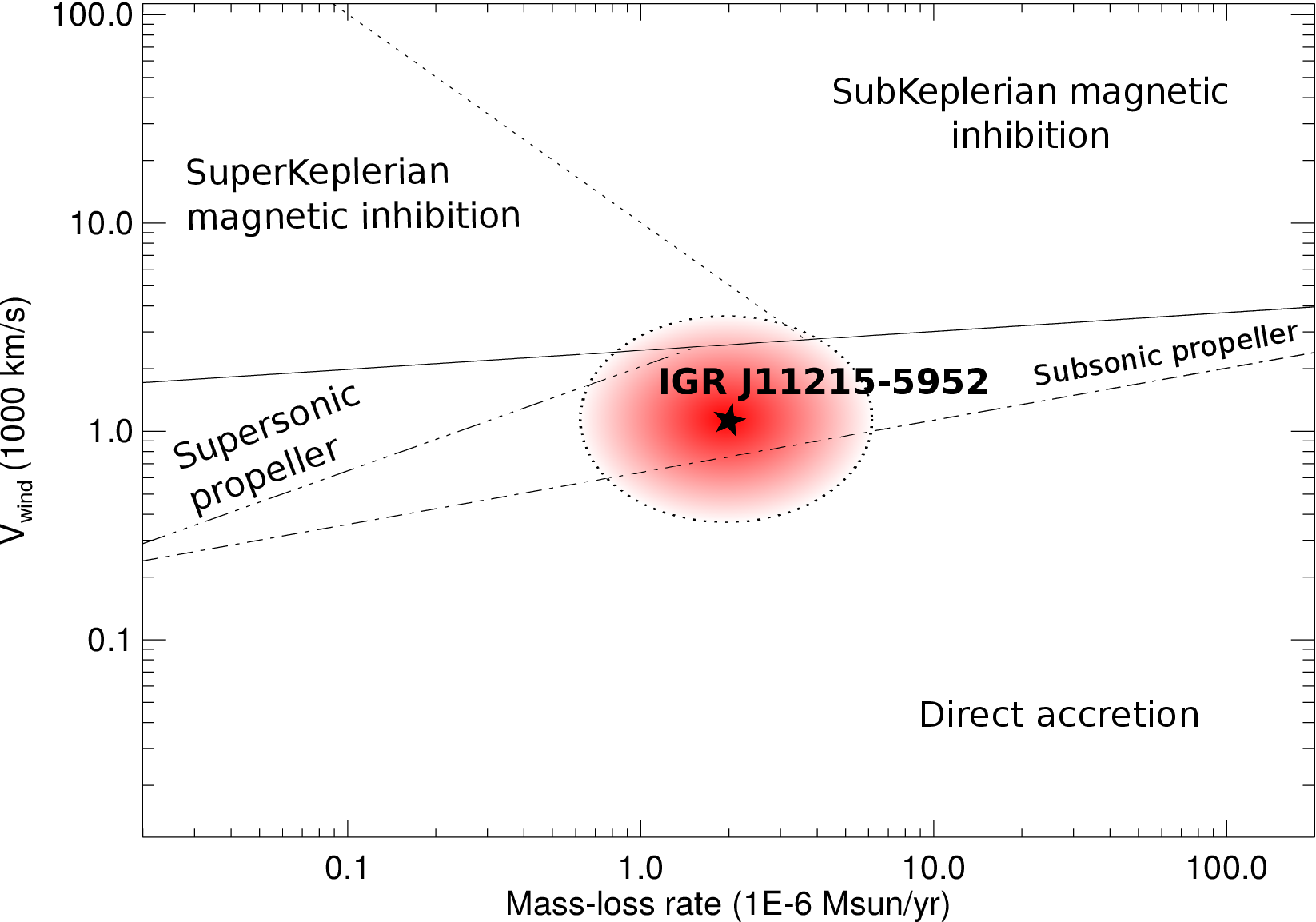, width=0.48\textwidth}
\epsfig{file=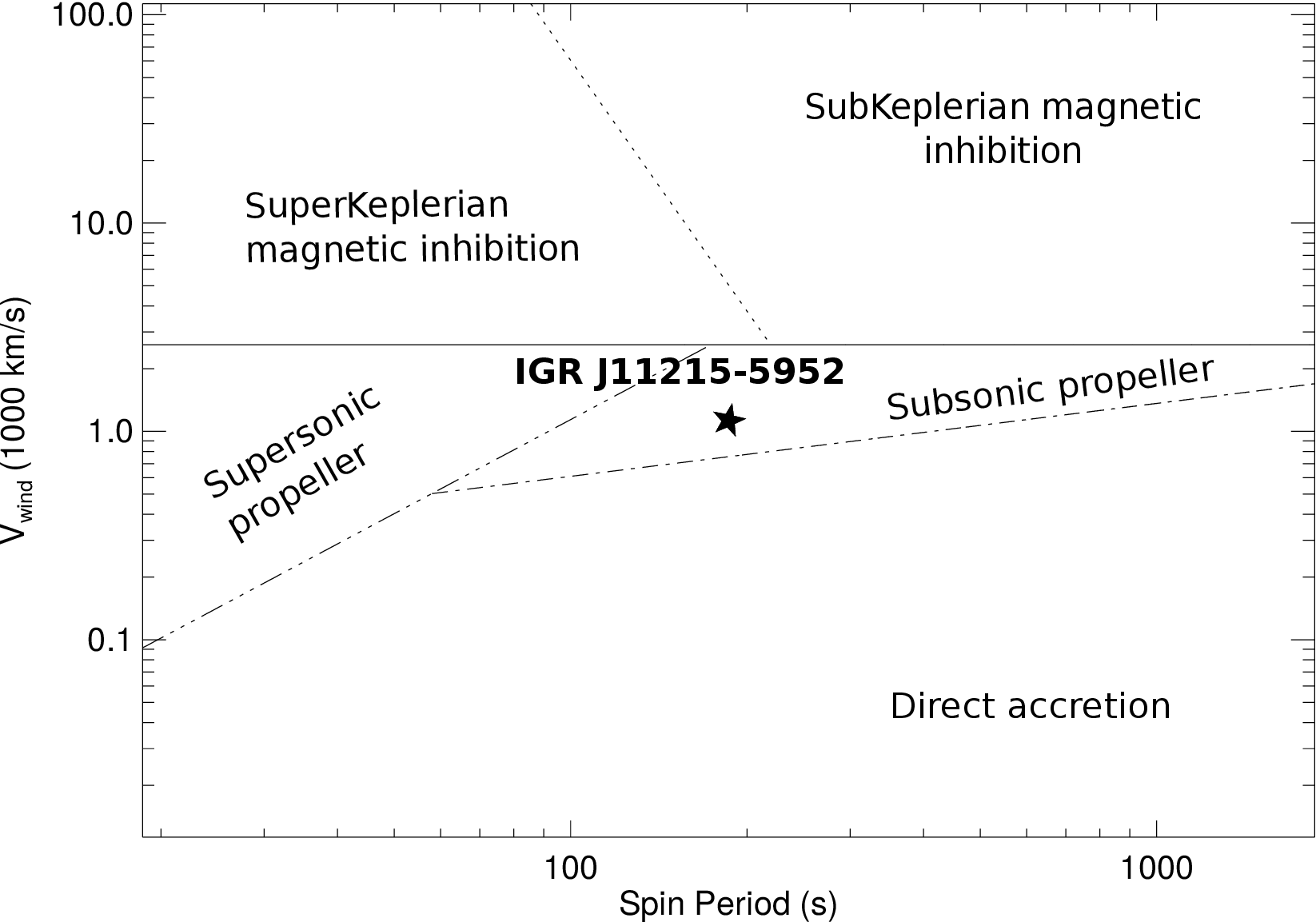, width=0.48\textwidth}
\caption{\footnotesize Left panel: position of IGR~J11215-5952 (SFXT) in the $\varv_\text{wind}$-$\dot{M}$ diagram. 
Right panel: position of IGR~J11215-5952 in the $\varv_\text{wind}$-$P_\text{spin}$ diagram. 
Equations 25, 26, 27 and 28 by \cite{2008ApJ...683.1031B} are represented by a solid, dotted, triple-dot-dashed and dot-dashed lines respectively.}
\label{fig:Bozzo_11215}
\end{figure}

\begin{figure}[H]
\centering
\epsfig{file=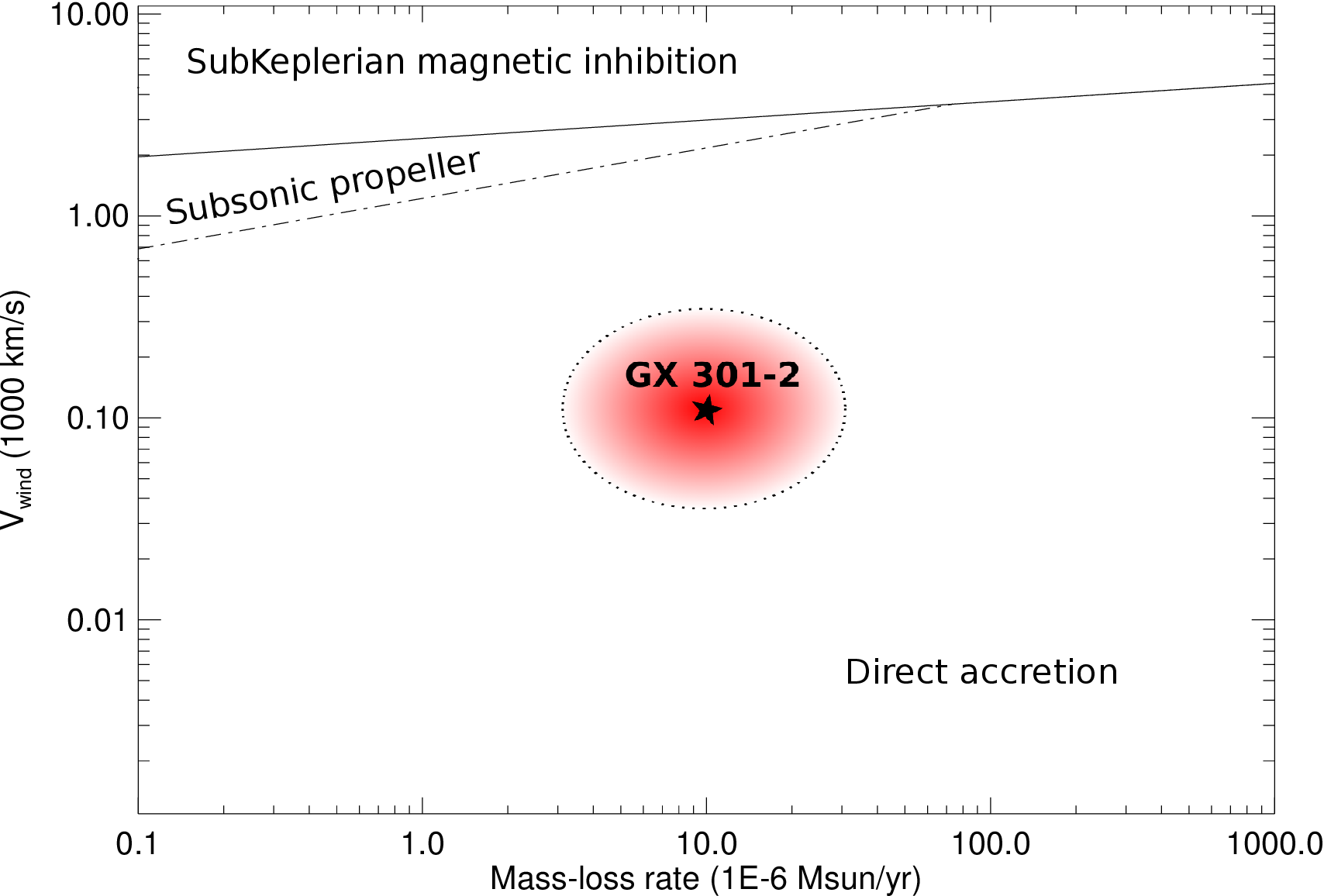, width=0.48\textwidth}
\epsfig{file=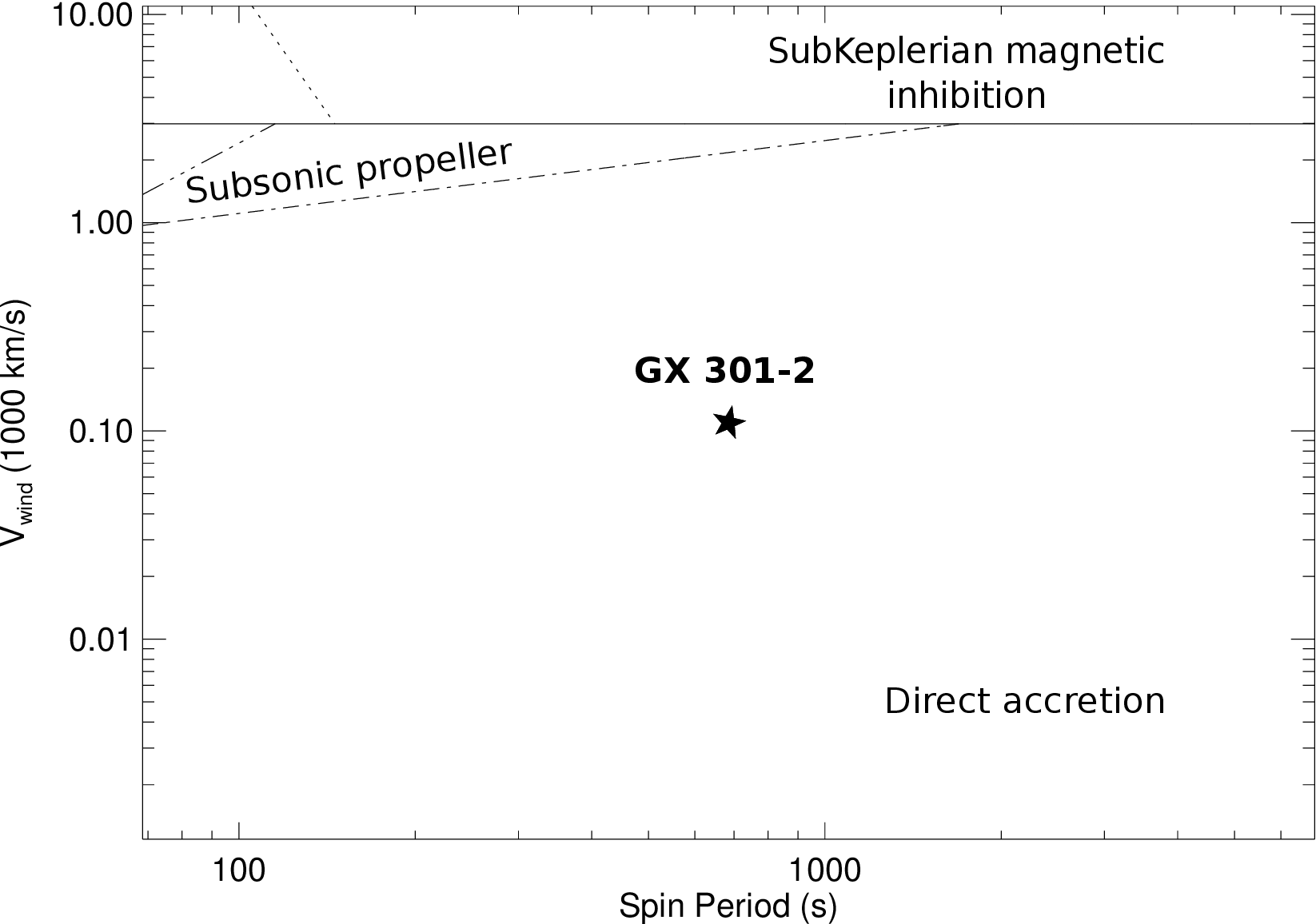, width=0.48\textwidth}
\caption{\footnotesize Left panel: position of GX~301-2 (SGXB) in the $\varv_\text{wind}$-$\dot{M}$ diagram. 
Right panel: position of GX~301-2 in the $\varv_\text{wind}$-$P_\text{spin}$ diagram.
Equations 25, 26, 27 and 28 by \cite{2008ApJ...683.1031B} are represented by a solid, dotted, triple-dot-dashed and dot-dashed lines respectively.} 
\label{fig:Bozzo_GX301}
\end{figure}

\begin{figure}[H]
\centering
\epsfig{file=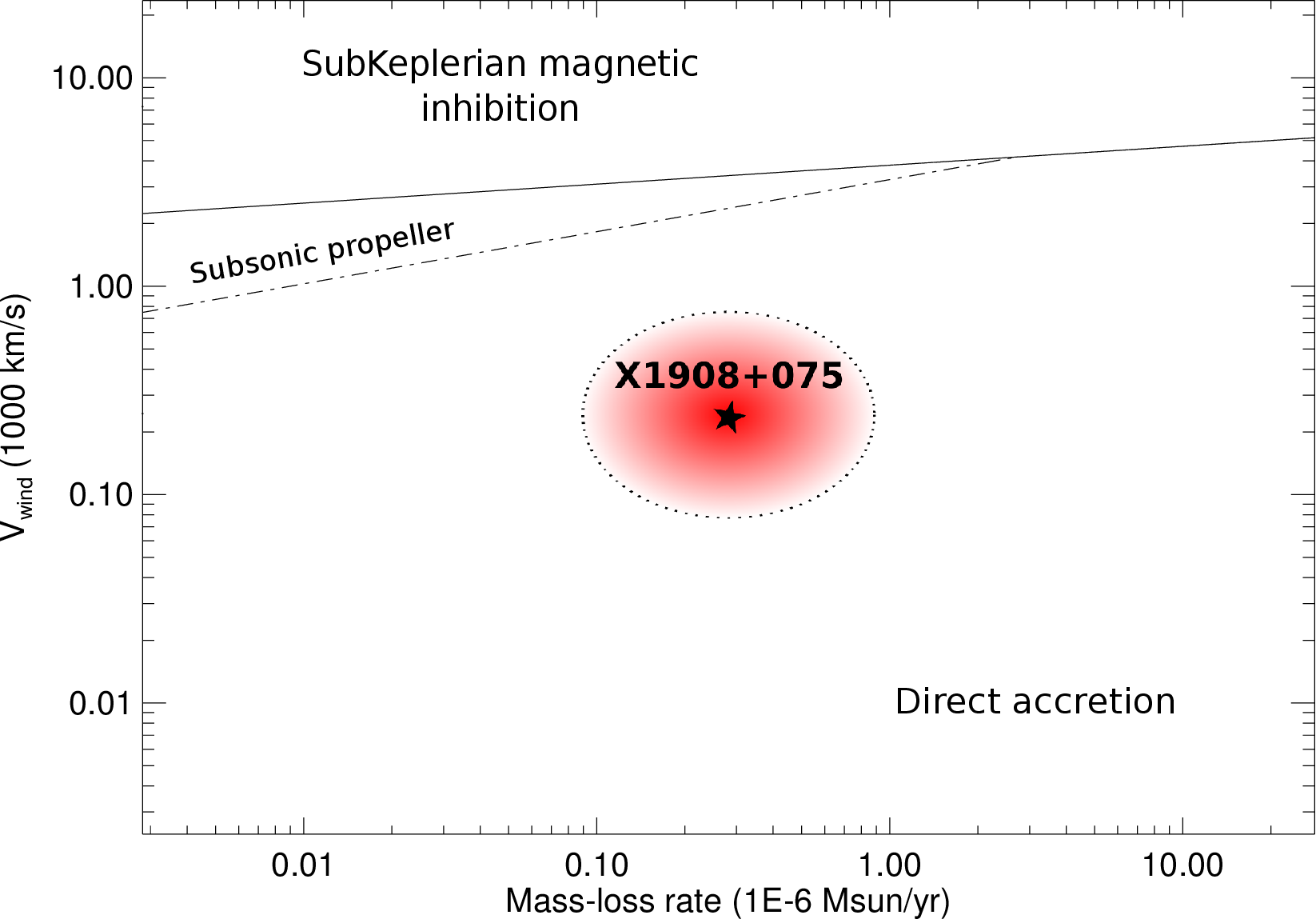, width=0.48\textwidth}
\epsfig{file=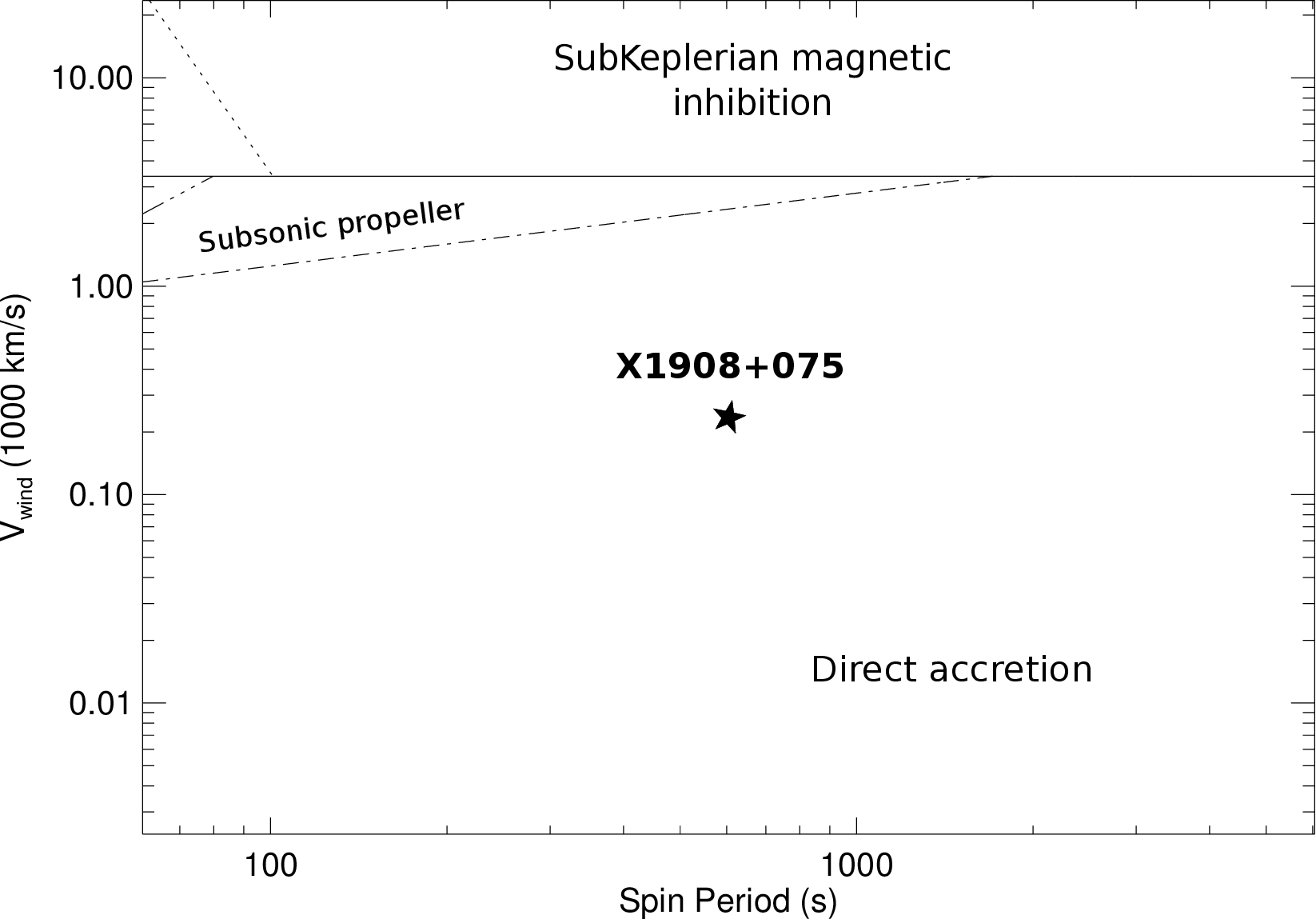, width=0.48\textwidth}
\caption{\footnotesize Left panel: position of X1908+075 (SGXB) in the $\varv_\text{wind}$-$\dot{M}$ diagram. 
Right panel: position of X1908+075 in the $\varv_\text{wind}$-$P_\text{spin}$ diagram.
Equations 25, 26, 27 and 28 by \cite{2008ApJ...683.1031B} are represented by a solid, dotted, triple-dot-dashed and dot-dashed lines respectively.} 
\label{fig:Bozzo_X1908}
\end{figure}

\begin{figure}[H]
\centering
\epsfig{file=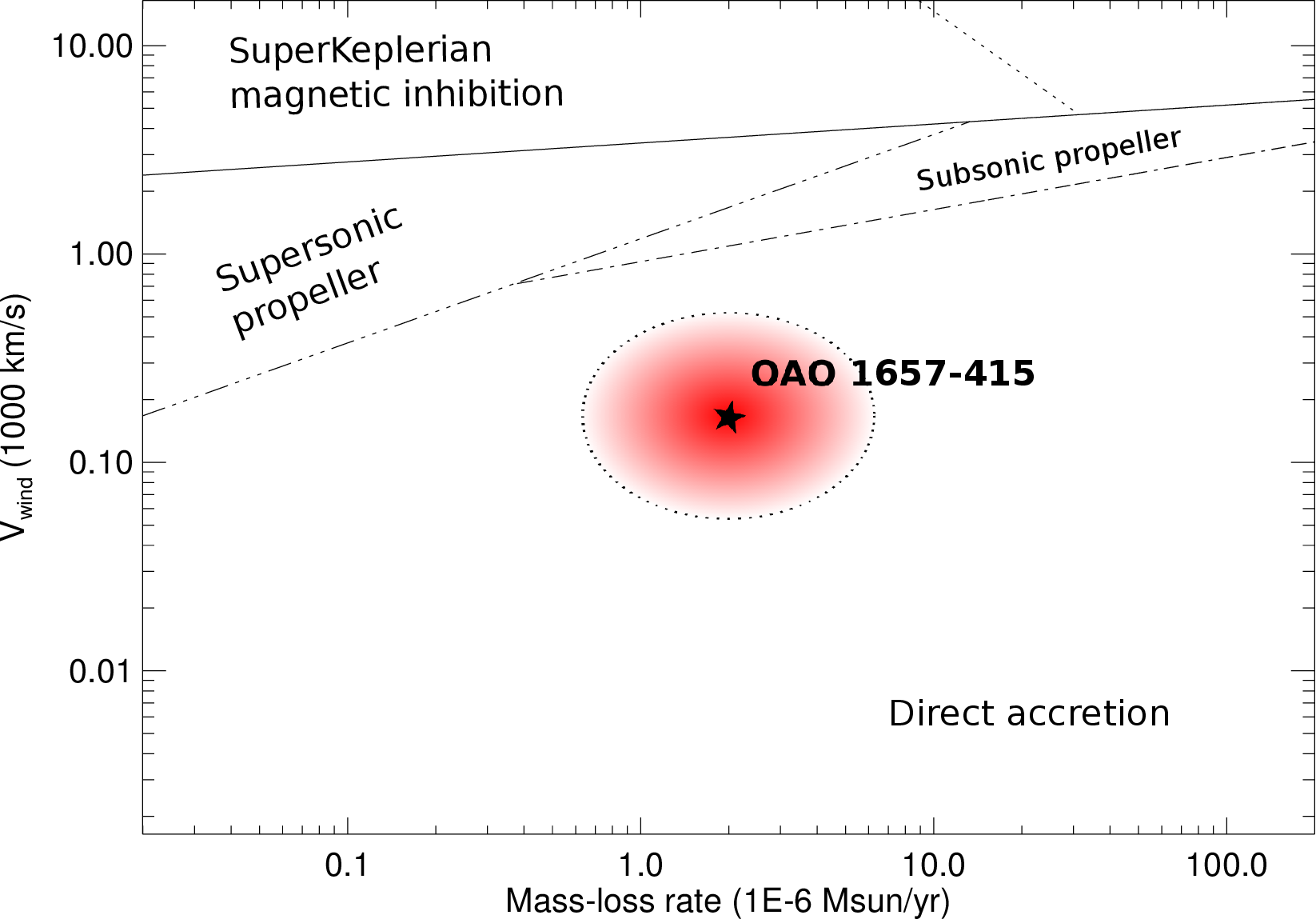, width=0.48\textwidth}
\epsfig{file=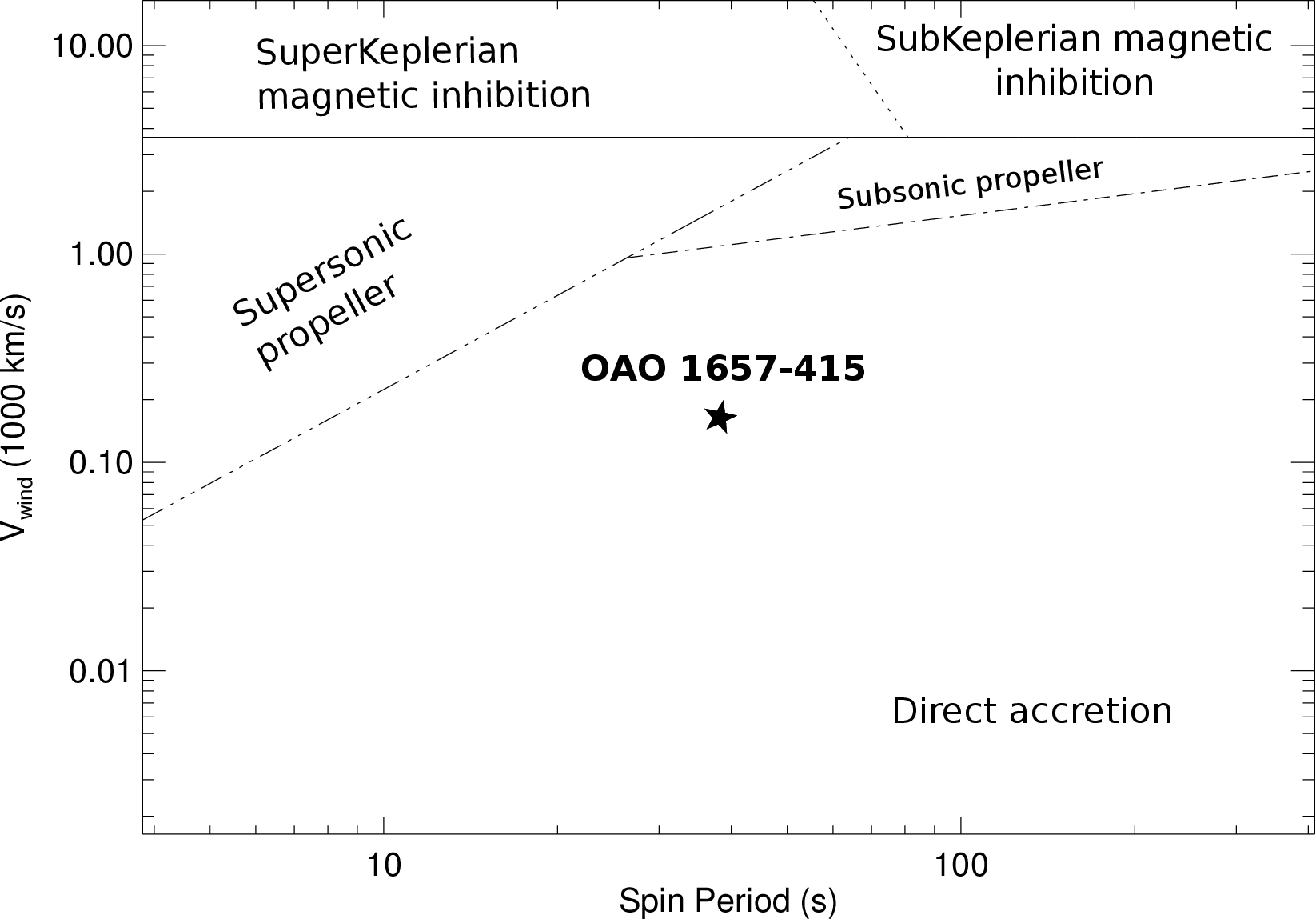, width=0.48\textwidth}
\caption{\footnotesize Left panel: position of OAO~1657-415 (SGXB) in the $\varv_\text{wind}$-$\dot{M}$ diagram. 
Right panel: position of OAO~1657-415 in the $\varv_\text{wind}$-$P_\text{spin}$ diagram. 
Equations 25, 26, 27 and 28 by \cite{2008ApJ...683.1031B} are represented by a solid, dotted, triple-dot-dashed and dot-dashed lines respectively.}
\label{fig:Bozzo_1657}
\end{figure}

\newpage
\section{Spectra \label{app:spectra}}
\subsection{IGR J17544-2619}
\begin{figure}[H]
\centering
\epsfig{file=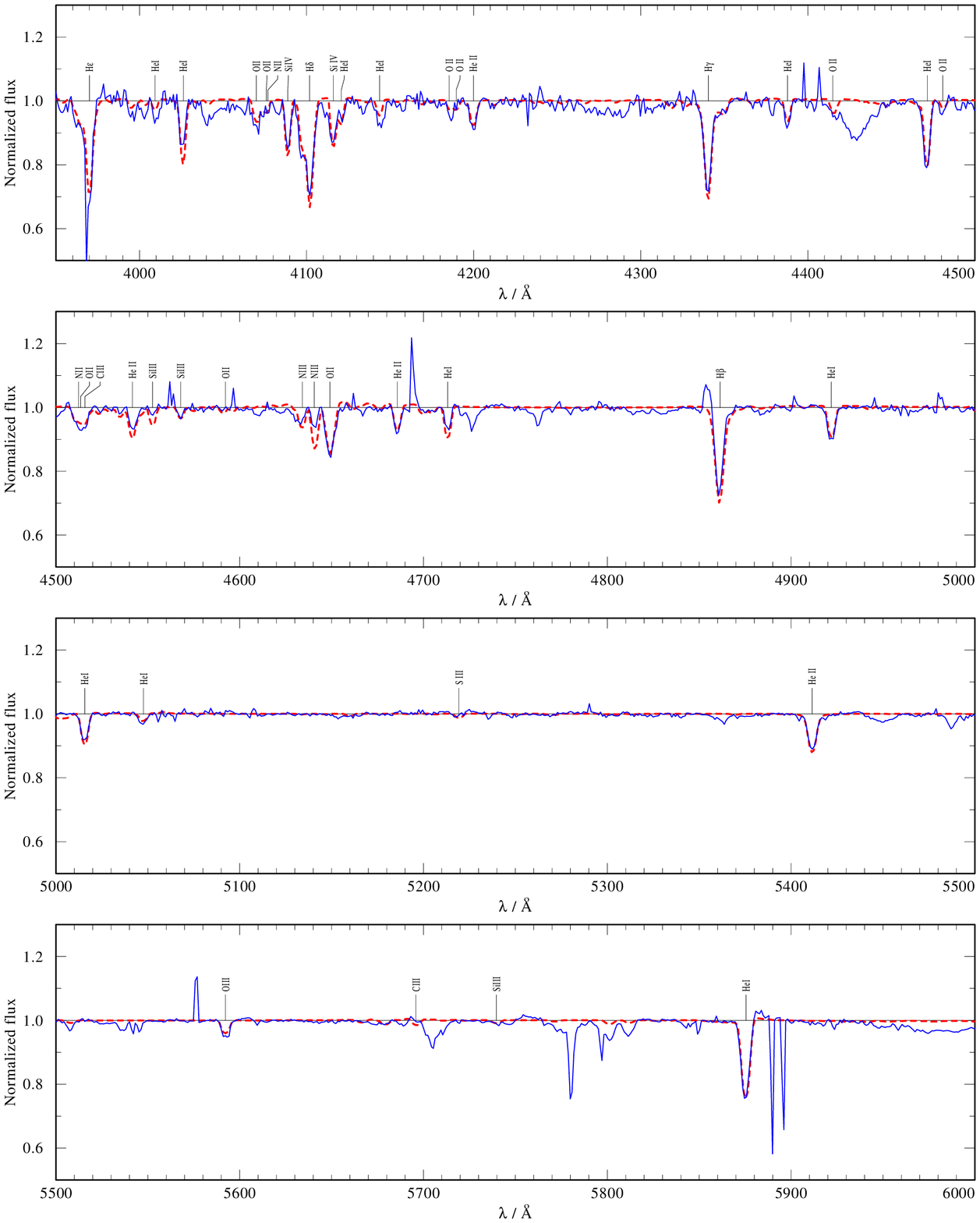, width=\textwidth}
\caption{\footnotesize continued.  }
\label{fig:summ_opt17544}
\end{figure}

\begin{figure}[H]
\centering
\ContinuedFloat
\epsfig{file=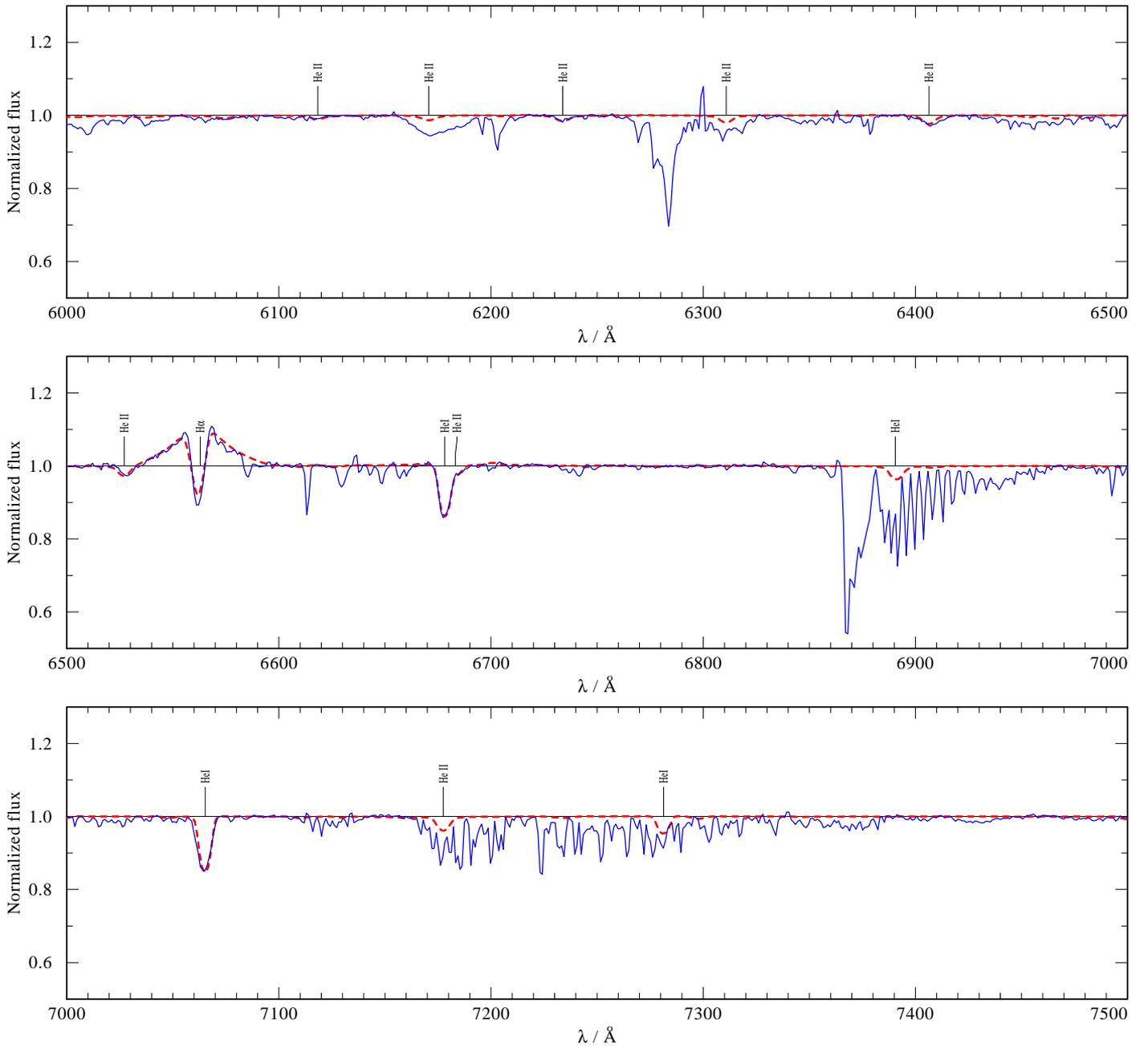, width=\textwidth}
\caption{\footnotesize Optical spectrum of IGR~J17544-2619 (blue), and the best 
fit model (red).  }
\label{fig:summ_opt217544}
\end{figure}

\begin{figure}[H]
\centering
\epsfig{file=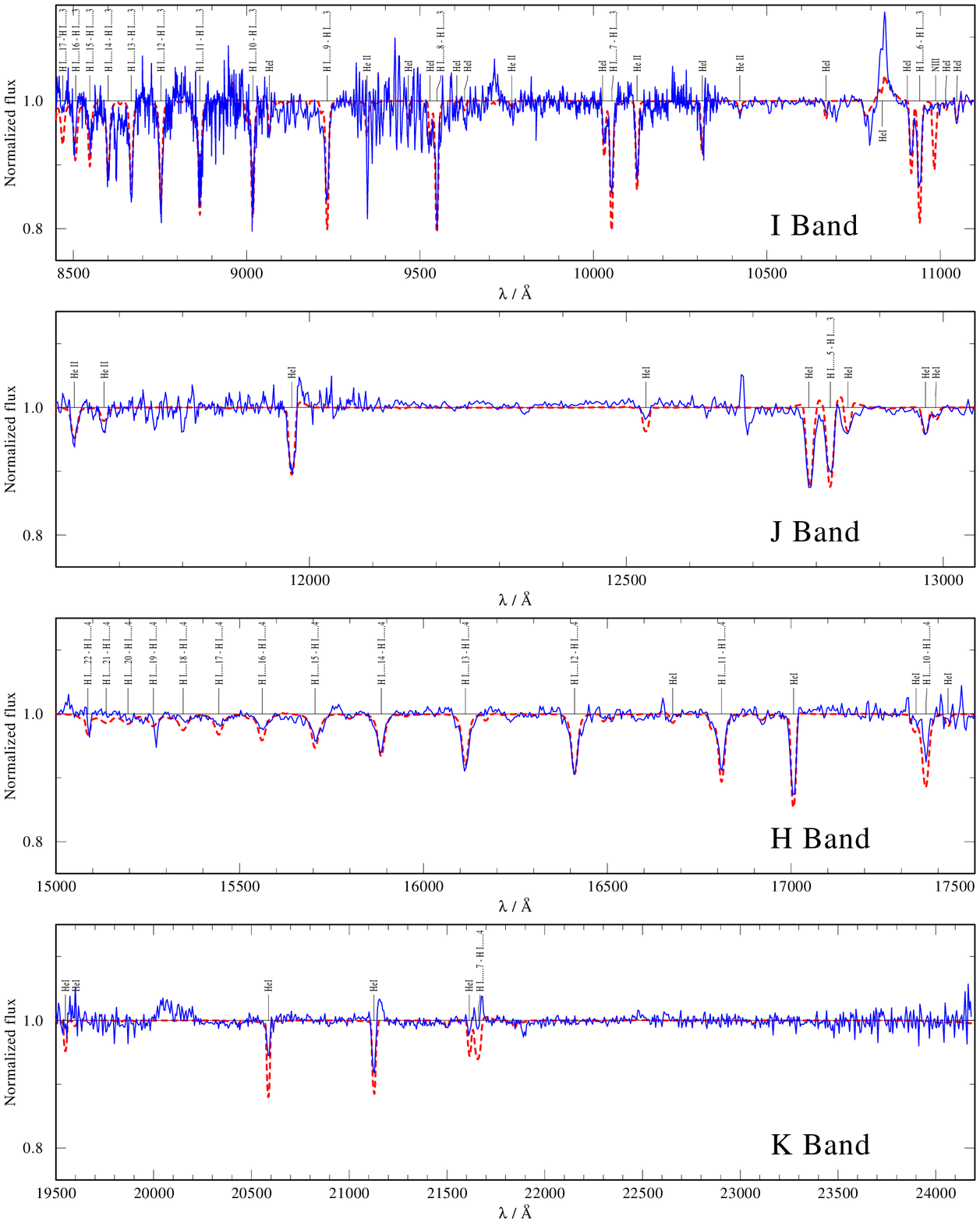, width=\textwidth}
\caption{\footnotesize Infrared spectrum of IGR~J17544-2619 (blue), and the best 
fit model (red).  }
\label{fig:summ_ir17544}
\end{figure}

\subsection{Vela X-1}
\begin{figure}[H]
\centering
\epsfig{file=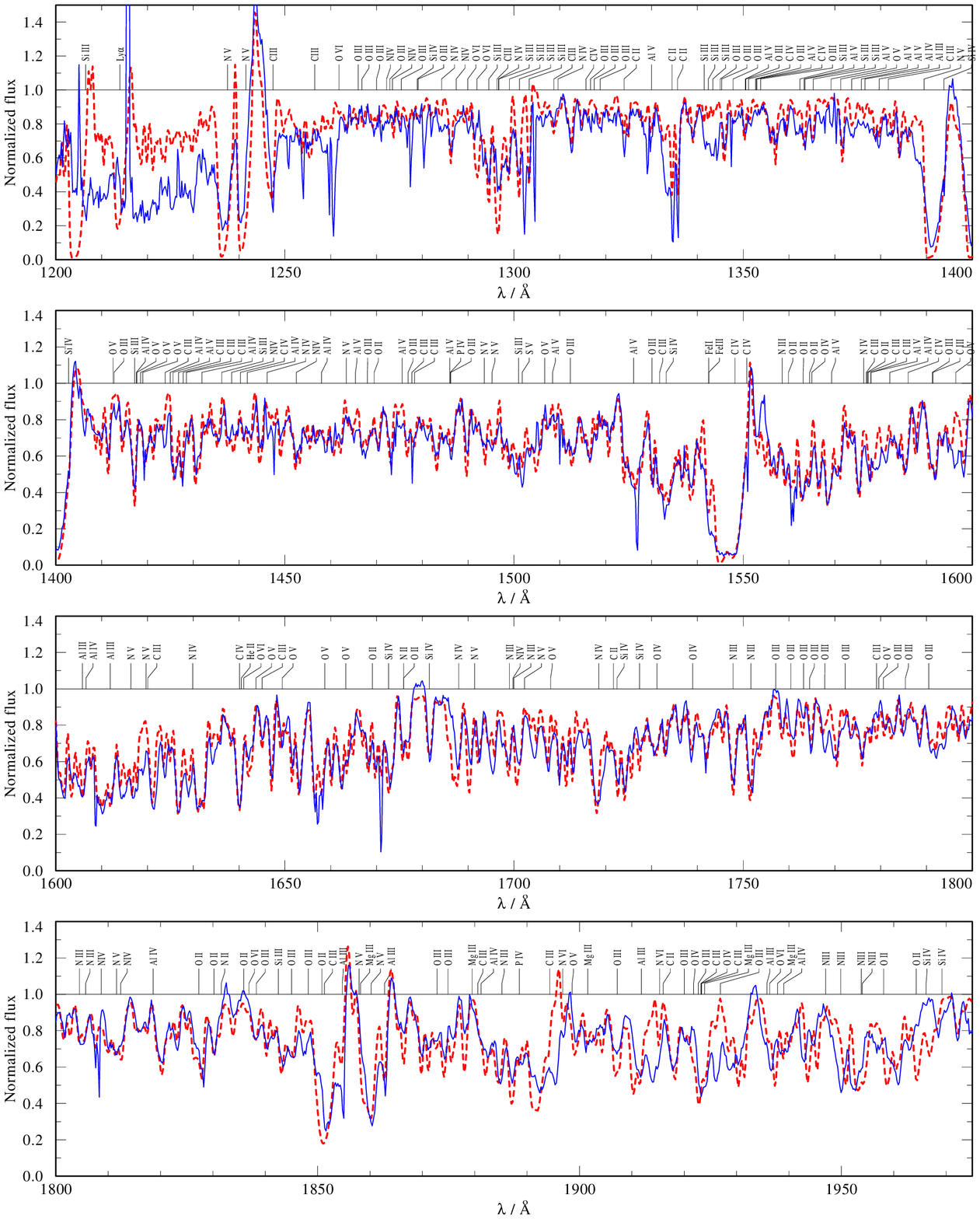, width=\textwidth}
\caption{\footnotesize Ultraviolet spectrum of Vela~X-1 (blue), and the best fit 
model (red).  }
\label{fig:summ_UVVela}
\end{figure}

\begin{figure}[H]
\centering
\epsfig{file=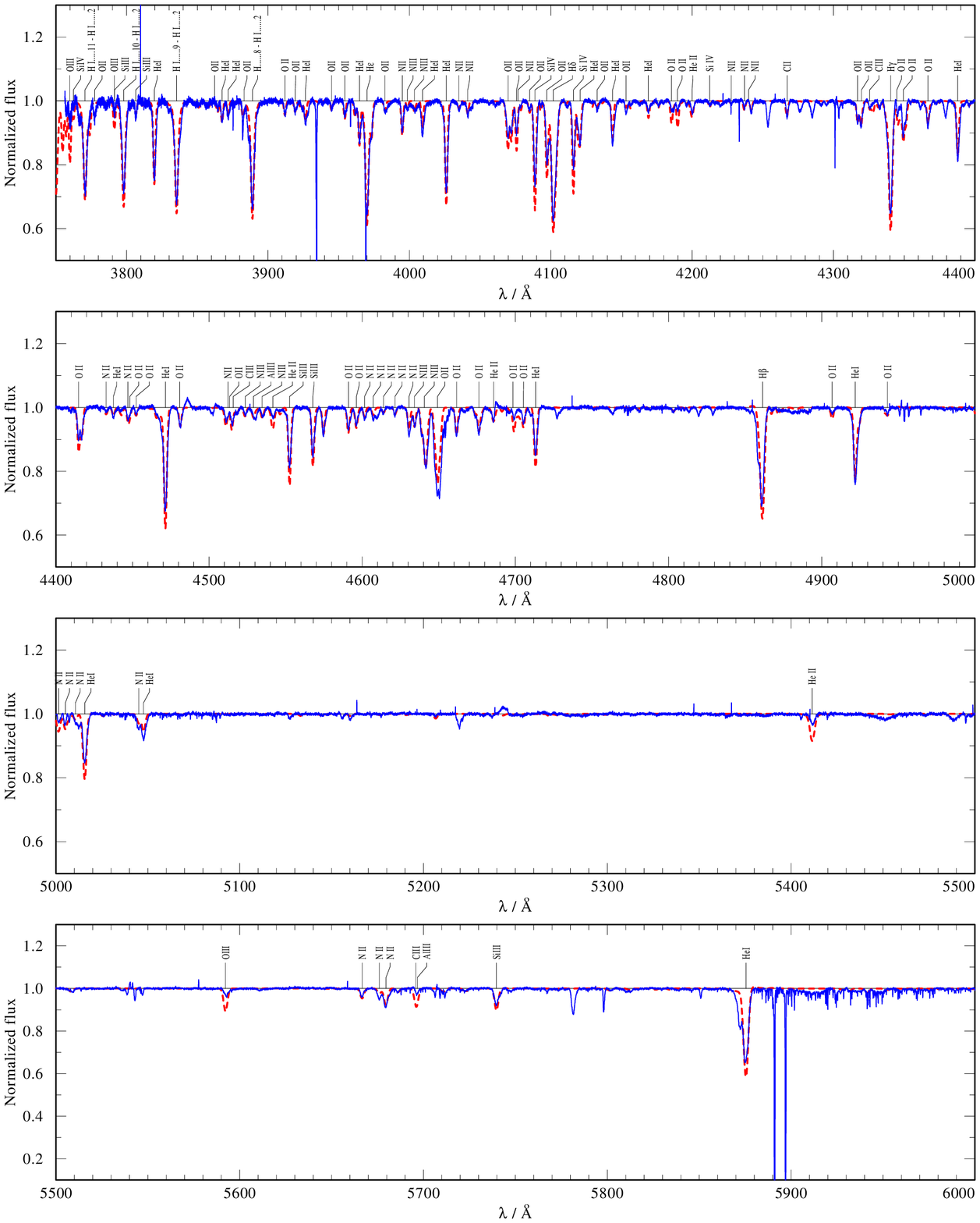, width=\textwidth}
\caption{\footnotesize continued.  }
\label{fig:summ_optVela}
\end{figure}

\begin{figure}[H]
\centering
\ContinuedFloat
\epsfig{file=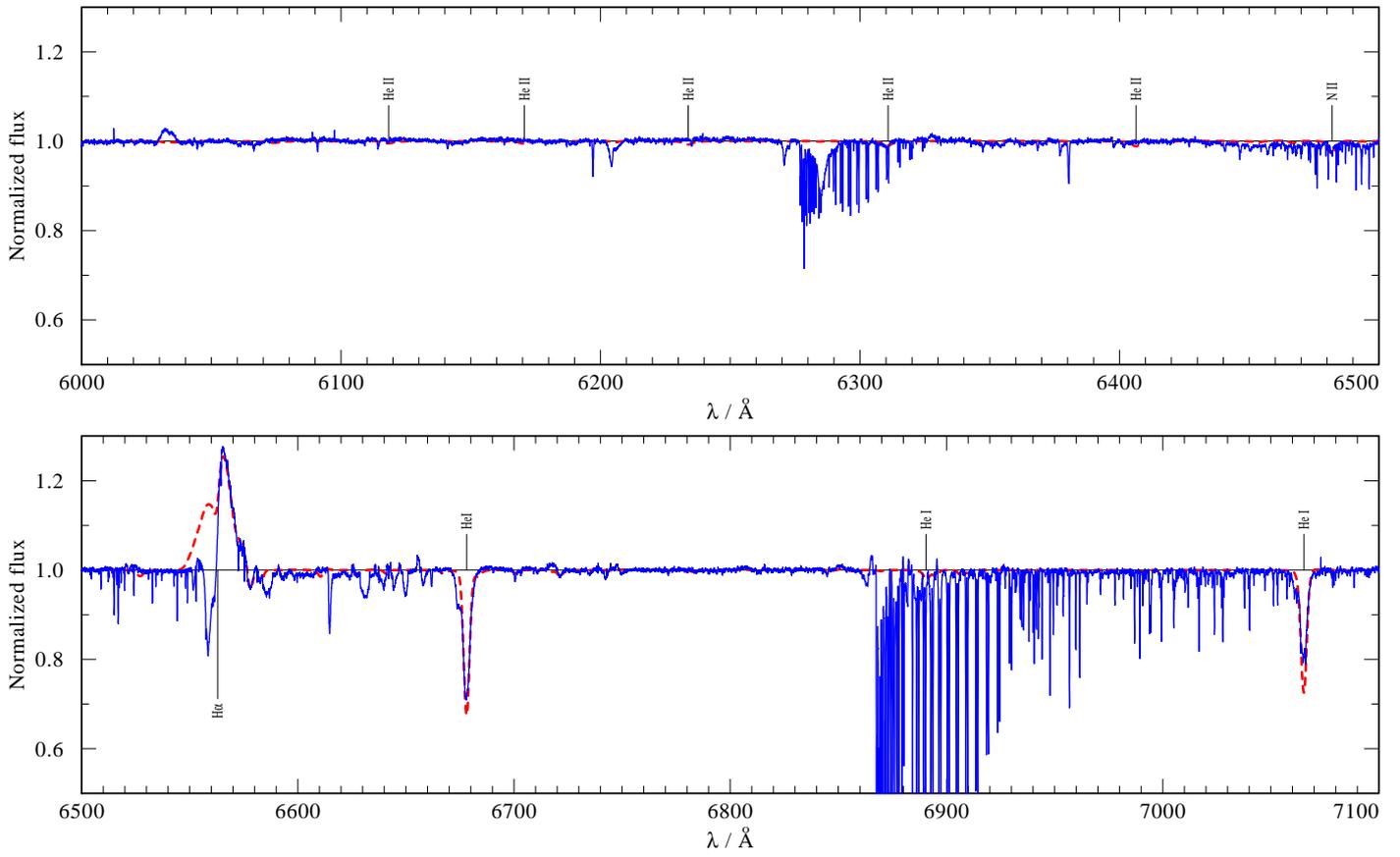, width=\textwidth}
\caption{\footnotesize Optical spectrum of Vela~X-1 (blue), and the best fit 
model (red).  }
\label{fig:summ_opt2Vela}
\end{figure}

\end{document}